\documentclass[prl,twocolumn,showpacs]{revtex4}
\usepackage{graphicx,amsfonts,amssymb,amsmath}

\begin{document}

\title{Capillary-Wave Description of Rapid Directional Solidification}

\date{\today}

\author{Alexander L. Korzhenevskii}
\affiliation{Institute for Problems of Mechanical Engineering,
RAS, Bol'shoi prosp. V. O., 61, St Petersburg, 199178, Russia}
\author{Richard Bausch}
\affiliation{Institut f{\"u}r Theoretische Physik IV,
 Heinrich-Heine-Universit{\"a}t D{\"u}sseldorf, Universit{\"a}tsstrasse
 1, D-40225 D{\"u}sseldorf, Germany}
\author{Rudi Schmitz}
\affiliation{Institut f{\"u}r Theoretische Physik A, RWTH Aachen University,
Templergraben 55, D-52056 Aachen, Germany}

\date{\today}

\begin{abstract}

A recently introduced capillary-wave description of binary-alloy solidification is generalized to include the procedure of directional solidification. For a class of model systems a universal dispersion relation of the unstable eigenmodes of a planar steady-state solidification front is derived, which readjusts previously known stability considerations. We, moreover, establish a differential equation for oscillatory motions of a planar interface that offers a limit-cycle scenario for the formation of solute bands, and, taking into account the Mullins-Sekerka instability, of banded structures.

\end{abstract}


\maketitle

\section{Introduction} 

The main feature of a recently introduced capillary-wave model \cite{PRE} for the solidification of a dilute binary alloy is the use of the interface position as a basic field variable, in addition to the concentration of the solute component. In the present paper this approach will be generalized to cover also the description of directional solidification, especially with regard to the rapid-growth regime. As outlined in reviews by Langer \cite{Langer-RMP} and by M\"uller-Krumbhaar et al. \cite{MKKB}, in directional solidification the growth of a crystal is accomplished by pulling it in opposite direction of an externally applied temperature gradient. We will mainly consider the case of a constant temperature gradient, which enters via a driving force in the equation of motion for the interface position. This form of description arises in the limit of an infinite heat conductivity from a more general model, involving energy density as an additional field variable. In general, such a model would allow to include the effect of heat diffusion.

As a first application of our approach we scrutinize  the possibility of stationary motions of a planar solidification front. The stability of such a front has been investigated in the rapid-growth regime with increasing regard of non-equilibrium effects by Mullins and Sekerka \cite{MS}, Coriell and Sekerka \cite{CS}, and by Merchant and Davis \cite{MD}. In Refs. \cite{CS} and \cite{MD} new oscillatory interface instabilities have been discovered in addition to the previously-known Mullins-Sekerka instability. Our own approach demonstrates that these effects are closely related to an instability, found by Cahn \cite{Cahn} in grain-boundary motion. The threshold of this instability represents a border line between regimes of steady-state and of non-steady-state motions of the solidification front.   

Non-steady interface motions operate in generating the periodic growth of layers with alternating homogeneous and dendritic micro-structures in binary alloys. This so-called banded structure occurs in many metallic alloys, as described in the review \cite{CGZK} by Carrard et al. who also offer a phenomenological explanation of the effect, using a quasi-stationary approximation. In a more microscopic treatment, Karma and Sarkissian \cite{KS1} pointed out that the banding phenomenon is due to relaxation oscillations of the solidification front. This behavior, described in more detail in Ref. \cite{KS2}, was derived from numerical solutions of the diffusion equations for the solute concentration in a dilute binary alloy and for the temperature, supported by non-equilibrium boundary conditions, as formulated by Aziz and Boettinger  \cite{AB}. Starting from a phase-field model, Conti has performed one- and two-dimensional simulations, describing the generation of solute bands \cite{C1}, and of banded structures \cite{C2}, respectively. He also has confirmed in Ref. \cite{C3} the observation by Karma and Sarkissian that the inclusion of heat diffusion leads to an increasing suppression of the formation of bands with decreasing heat conductivity.

In the application of our capillary-wave description we are going to analyze the simplest-possible model, which shows the banding effect. We, accordingly, consider the dynamics of a planar interface, neglecting heat diffusion, and assuming an overall constant diffusion coefficient for the solute component. For a class of model systems with arbitrary equilibrium profiles of the solute concentration these properties lead to an integro-differential equation for the interface position. In case of a sufficiently small temperature gradient, realized in many experiments, this equation can be reduced to the differential equation of a damped nonlinear oscillator \cite{PRL}. Stationary solutions of this equation turn out to exist in some region, limited by the threshold of the Cahn instability. This instability is attended by an oscillation, blowing up in the unstable regime to a limit-cycle behavior of the solidification front and of the solute concentration at the interface. Close to the stability threshold the transition from uniform to periodic solutions can analytically be evaluated by the Bogoliubov-Mitropolsky method \cite{Bogoliubov}. The nature of the periodic solutions can be tuned from almost-harmonic to distinctive relaxation oscillations by changing the pulling velocity from the stability threshold to values deep inside the unstable regime.

A benefit of our approach is that, apart from solving the oscillator equation in the unstable regime, all steps of the procedure could be accomplished analytically, which deepens our understanding of the banding effect. The discussion of micro-segregation effects at an oscillating solidification front requires to consider the stability of such a front in the transverse direction. Since, however, our results reveal an almost stationary behavior of the interface motion in the so far barely understood low-velocity regime \cite{KS2}, we presently only complement the established limit-cycle scenario by the standard Mullins-Sekerka procedure. Then, in some window of the model parameters, the high- and low-velocity sections of a limit cycle are located inside the Mullins-Sekerka stable and unstable regimes. As a result, a dendritic microstructure will develop in the low-velocity bands, which we consider as a kind of noise on the more macroscopic scale of the periodic array of the widely flat bands. 

\section{Capillary-Wave Model} 

The effective Hamiltonian of our capillary-wave model is a functional of the interface position $Z({\bf x},t)$ and of the excess concentration $C({\bf r},t)$ of the solute relative to its value $C_S$ in the solid phase. In terms of these field variables the effective Hamiltonian has the form
\begin{eqnarray}\label{I-Hamiltonian}
H&=&\frac{\sigma}{2}\int d^2x\,(\partial Z)^2\nonumber\\
&+&\frac{\kappa}{2}\int d^3r\,\Bigl[C-U(z-Z)\Bigr]^2\,\,,
\end{eqnarray}
established already in Ref. \cite{PRE}. It determines all static properties of the system in thermal equilibrium at some fixed temperature $T_S<T_M$ where $T_M$ denotes the melting temperature of the solvent, showing up in the temperature-concentration phase diagram, Fig. 1. The input quantities in the Hamiltonian (\ref{I-Hamiltonian}) are the surface tension $\sigma$, the coupling parameter

\begin{equation}\label{kappa}
\kappa=-\left(\frac{\partial C_L}{\partial T}\right)^{-1}\frac{L}{T_M}\,\frac{1}{\Delta C}\,\,,
\end{equation}

\begin{center}
\begin{figure}[t]
 \includegraphics[width=8cm]{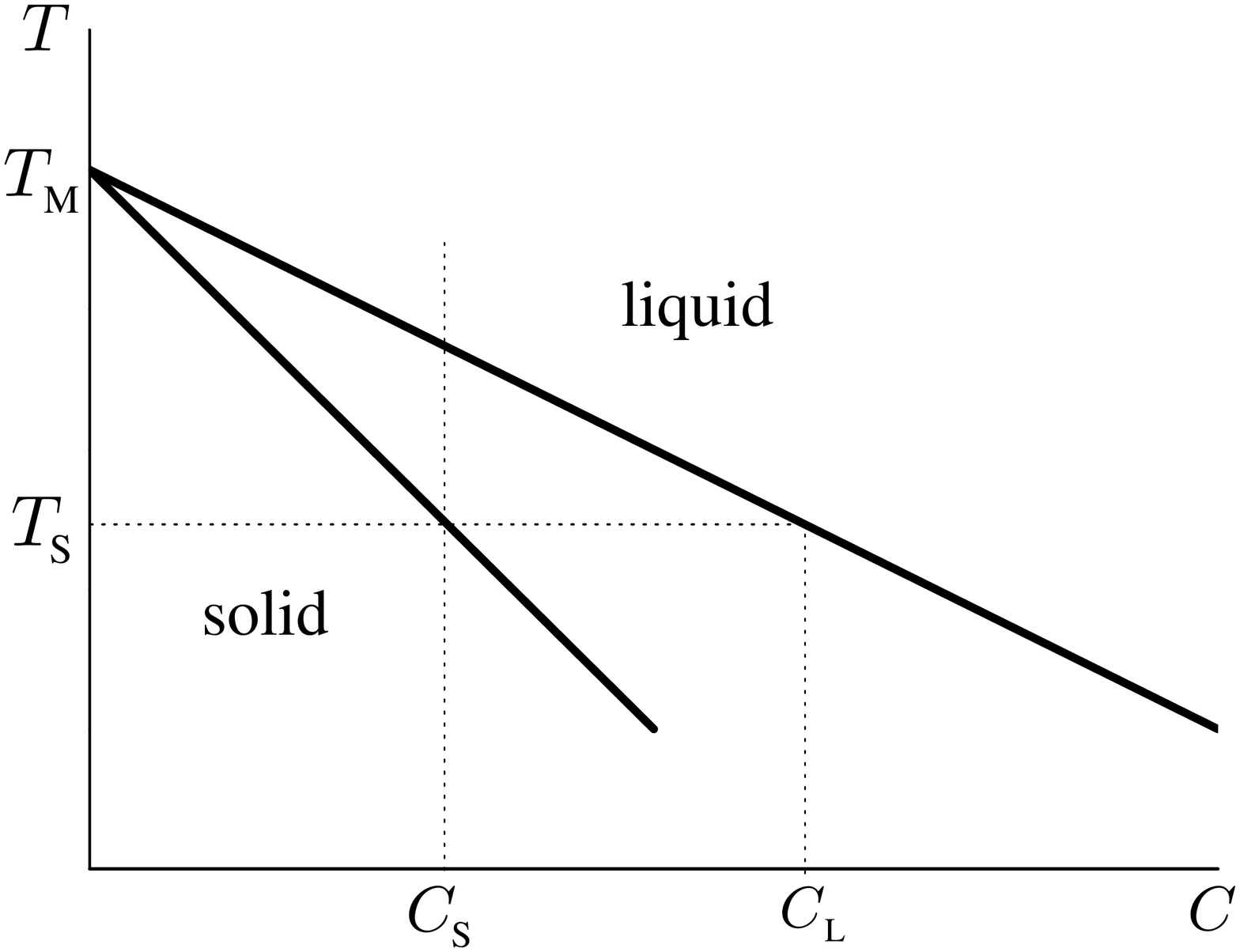}
 \caption{Temperature-concentration phase diagram, showing the liquidus and solidus lines $T_L(C),T_S(C)$ which meet at $T_M$. The values $C_L$ and $C_S$ refer to the temperature $T_S$.}
 \end{figure}
 \end{center}
involving the solute concentration $C_L$ in the liquid phase, the latent heat $L$ per unit volume, the miscibility gap

\begin{equation}\label{misc-gap}
\Delta C\equiv C_L-C_S\,\,,
\end{equation}
visible in Fig. 1, and the solute-concentration profile in thermal equilibrium,

\begin{equation}\label{C_E-U}
U(z-Z)=C_E(z-Z)\,\,.
\end{equation}
Whereas the expression (\ref{kappa}) has been derived in Ref. \cite{PRE}, Eq. (\ref{C_E-U}) directly follows from the equilibrium condition $\delta H/\delta C=0$.

From Ref. \cite{PRE} we also adopt the equations of motion

\begin{eqnarray}\label{I-dynamics}
\partial_t Z&=&\Lambda\,\left(F-\frac{\delta H}{\delta Z}\right)\,\,,\nonumber\\
\partial_t C&=&D\,\nabla^2\frac{1}{\kappa}\,\frac{\delta H}{\delta C}\,\,
\end{eqnarray}
where the rate $\Lambda$ measures the interface mobility, and $D$ is the diffusion coefficient of the solute, here assumed to have an overall constant value. The externally applied temperature gradient $S$, and the pulling velocity $V_P$ enter via the driving force

\begin{equation}\label{force}
F=L\,\frac{T_S-T}{T_M}
\end{equation}
where, in terms of the temperature $T_P$ at the steady-state position $Z(t)=V_P\,t$,

\begin{equation}\label{int-temp}
T=T_P+S(Z-Z_P)\,\,\,\,,\,\,\,Z_P=V_P\,t\,\,.
\end{equation}
In Appendix A we will consider a more general model, which includes energy density as an additional field. We also will show that the reduced model equations (\ref{I-Hamiltonian}) - (\ref{int-temp}) emerge in the limit of an infinite heat conductivity.

A dimensionless form of the model equations (\ref{I-dynamics}) can be obtained by adopting from Ref. \cite{PRE} the mappings

\begin{eqnarray}\label{rescaling}
&&\frac{1}{\xi}\,{\bf r}\rightarrow{\bf r}\,\,\,,\,\,\,\frac{D}{\xi^2}\,t\rightarrow t\,\,\,,\,\,\, \frac{\xi}{\sigma}\,F\rightarrow F\,\,,\nonumber\\&& \frac{2}{\Delta C}\,C\rightarrow C\,\,\,,\,\,\,\frac{2}{\Delta C}\,U\rightarrow U\,\,,
\end{eqnarray}
where, in the present context, the length $\xi$ is defined by 

\begin{equation}\label{length}
\xi\equiv\frac{\Delta C}{2}\frac{1}{U'(0)}\,\,.
\end{equation}
In terms of the dimensionless quantities

\begin{eqnarray}\label{dimensionless}
&&\gamma\equiv\frac{\xi\kappa}{\sigma}\left(\frac{\Delta C}{2}\right)^2\,\,\,,\,\,\, p\equiv \frac{\Lambda\sigma}{D} \,\,,\nonumber\\&&m^2\equiv\frac{\xi^2 L}{\sigma}\,\frac{S}{T_M}\,\,,
\end{eqnarray}
the resulting equations of motion read

\begin{eqnarray}\label{scal-eq}
&&\frac{1}{p}\,\partial_t Z=F(Z)-\gamma\int_{-\infty}^{+\infty}dz\,U'(z-Z)[C-U(z-Z)] \,\,,\nonumber\\&&\partial_t C=D\,\nabla^2[C-U(z-Z)])\,\,,
\end{eqnarray}
with the driving force $F$ given by

\begin{equation}\label{scaled-F}
F=F_P-m^2(Z-Z_P)\,\,\,,\,\,F_P\equiv\frac{\xi L}{\sigma}\,\frac{T_S-T_P}{T_M}\,\,.
\end{equation}
As a first application of Eqs. (\ref{scal-eq}) and (\ref{scaled-F}) we now will consider the steady-state motion of a planar interface with velocity $V_P$ in $z$-direction.

\section{Stationary Planar Growth}

Measuring velocities in units of the diffusion velocity,

\begin{equation}\label{velocities}
v\equiv \frac{V}{V_D}\,\,\,,\,\,\,V_D\equiv\frac{D}{\xi}\,\,,
\end{equation}
the stationary growth of a planar solidification front is described in the co-moving frame $z=v_P\,t+\zeta$ by the equations

\begin{eqnarray}\label{balance}
&&\frac{1}{p}\,v_P=F_P+G_P(v_P)-G_P(0)\,\,,\\&&G_P(v_P)\equiv-\,\gamma\int_{-\infty}^{+\infty}d\zeta\,
U'(\zeta)\,C_P(\zeta;v_P)\,\,,\nonumber\\&&C_P(\zeta;v_P)=\int_{-\infty}^{\zeta}d\zeta'\,U'(\zeta')\,
\exp{[v_P(\zeta'-\zeta)]}\,\,.\nonumber
\end{eqnarray}
These equations are identical to those, derived in Ref. \cite{PRE} for the case of solidification by under-cooling the liquid phase from $T_S$ to $T_P$. In particular, the result (\ref{balance}) is independent of the parameter $m^2$, which only enters in discussing the stability of the planar morphology.

For perturbations of the form

\begin{eqnarray}\label{perturbations}
&&h({\bf x},t)\equiv Z({\bf x},t)-v_Pt\,\,,\\
&&c({\bf x},\zeta,t)\equiv C({\bf x},\zeta,t)-C_P(\zeta;v_P)+C_P'(\zeta;v_P)h({\bf x},t)\,\,,\nonumber
\end{eqnarray}
the resulting equations of motion read

\begin{eqnarray}\label{increments}
&&\frac{1}{p}\,\partial_t h=(\partial^2-m^2)h-\int_{-\infty}^{+\infty}d\zeta\,U'(\zeta)\,c({\bf x},\zeta,t)\,\,,\\&& \partial_t c=v_P\,\partial_\zeta c+(\partial_\zeta^2+\partial^2)\,c+ [C_P'(\partial_t-\partial^2) +U'\partial^2]\,h\nonumber
\end{eqnarray}
where $\partial^2\equiv\nabla^2-\partial_\zeta^2$. These equations have eigensolutions of the form

\begin{eqnarray}\label{monochrom}
&&h({\bf x},t)=\hat h({\bf q},\omega)\exp{(i{\bf q}\cdot{\bf x}+\omega t)}\,\,,\\
&&c(\zeta,{\bf x},t)=\hat c(\zeta,{\bf q},\omega)\exp{(i{\bf q}\cdot{\bf x}+\omega t)}\,\,,\nonumber
\end{eqnarray}
which, after elimination of the component $\hat c$, lead to the eigenmode dispersion relation

\begin{eqnarray}\label{dispersion}
\frac{\omega}{p}+q^2+m^2-v_P[G_P(v_P+\lambda)-G_P(v_P)]&=&\nonumber\\\frac{\lambda^2-q^2 }{v_P+2\lambda}\, [G_P(v_P+\lambda)+ G_P(\lambda)]
\end{eqnarray}
where, deriving from the equation of motion for $c$, 

\begin{equation}\label{root}
\lambda\equiv -(v_P/2)+\sqrt{(v_P/2)^2+\omega+q^2}\,\,.
\end{equation}
The result (\ref{dispersion}) is similar to that, established in Ref. \cite{PRE} and only differs by the additional term $m^2$.

Inspection of the low\,-\,\,$q,\omega$ behavior of the dispersion relation (\ref{dispersion}) leads to identify an eigenfrequency $\omega_1(q)$, which captures the Mullins-Sekerka instability. Since the parameter $m$ acts as a long-wavelength cutoff, the wave-number threshold $q_c$ for this instability is shifted from $q_c=0$ at $m=0$ to some finite value, determined by the relations $\omega_1(q_c)=\omega_1'(q_c)=0$. Elimination of $q_c$ then generates the neutral stability curve of the instability in form of a function $v_P(\gamma)$, with a parametric dependence on $m$. An explicit form of this neutral line will later be derived for a specific expression of $U(z-Z)$. 

A second branch $\omega_2(q)$ comprises an instability, similar to that, discovered by Cahn \cite{Cahn} in the process of grain-boundary motion. This instability is characterized by a gap at $q=0$, determined by the relation

\begin{equation}\label{gap}
\frac{\omega_2(0)}{\Omega(v_P)}=-\,\frac{F_P'(v_P)\Omega(v_P)}{2m^2}\pm\sqrt{\frac{[F_P'(v_P)\Omega(v_P)]^2}
{4m^4}-1}\,\,,\nonumber
\end{equation}

\begin{equation}\label{pulsatile}
\Omega(v_P)\equiv 
m\left\{\frac{d^{\,2}}{dv_P^{\,2}}\left[-\,\frac{G_P(v_P)+G_P(0)}{2\,v_P}\right]\right\}^{-1/2}\,\,.
\end{equation}
In the limit $m\rightarrow 0$ a single nonzero value of the gap survives, which is identical to that, found in Ref. \cite{PRE}. For $m\ne 0$ the neutral stability curve $F_P'(v_P)=0$ of the Cahn instability is attended by an oscillation of period $\Omega(v_P)$. Similar oscillatory instabilities have previously been observed by Coriell and Sekerka \cite{CS}, and later by Merchant and Davis \cite{MD}. However, the neutral line, found in Ref. \cite{MD}, differs from ours, which, we conjecture, arises from an unsettled generalization of the Gibbs-Thomson relation. We next will demonstrate that the instability, described by Eq. (\ref{gap}), acts as a seed for the limit-cycle behavior in the unstable regime.

\section{Non-Stationary Planar Growth}

For general unsteady motions of a planar interface the equations of motion (\ref{scal-eq}) reduce to the form

\begin{eqnarray}\label{integro}
&&\frac{1}{p}\,\dot Z(t)=F_P-m^2[Z(t)-Z_P(t)]\nonumber\\ &&-\,\gamma\int_{-\infty}^{+\infty}dz\,U'(z-Z(t))[C(z,t)-U(z-Z(t))]\,\,,\nonumber\\\,\nonumber\\
&&(\partial_t-\partial_z^2)C(z,t)=-\,U''(z-Z(t))\,\,.
\end{eqnarray}
We are mainly interested in the late-stage behavior of $Z(t)$, and, therefore, are going to replace in the first of the equations (\ref{integro}) the solution $C(z,t)$ of the second equation, complemented by the  boundary condition $C(z,-\infty)=0$. This leads to the expression

\begin{eqnarray}\label{concentration}
C(z,t)=\int_{-\infty}^{t}dt'\int_{-\infty}^{+\infty}dz'\,\partial_{z'}\mathcal{G}(z-z',t-t')&&\nonumber\\
\cdot\,U'(z'-Z(t'))\,\,,&&
\end{eqnarray}
involving the Green function 

\begin{equation}\label{Green}
\mathcal{G}(z,t)=\int_{-\infty}^{+\infty}\frac{dk}{2\pi}\,\exp{(-k^2t+ik\,z)}\,\,.
\end{equation}

After the variable substitutions

\begin{equation}\label{substitutions}
\zeta\equiv z-Z(t)\,\,\,,\,\,\,\zeta'\equiv z'-Z(t')\,\,,
\end{equation}
and expansion of $Z(t')$ around $Z(t)$, we obtain

\begin{eqnarray}\label{expansion}
&&\mathcal{G}(\zeta-\zeta'+Z(t)-Z(t'),\,t-t')=\\&&\,\nonumber\\
&&\int_{-\infty}^{+\infty}\frac{dk}{2\pi}\exp{[-k^2(t-t')+ik(\zeta-\zeta')+ik(t-t')v(t)]}\nonumber\\
&&\cdot\,\exp{\left[-ik\sum_{n\ge 2}\frac{(-1)^n}{n!}(t-t')^n\partial_t^{\,n-1}v(t)\right]}\,\,,\nonumber
\end{eqnarray}
using the notation

\begin{equation}\label{velocity} 
v(t)\equiv\dot Z(t)=v_P+\dot h(t)\,\,. 
\end{equation}

For $m^2\ll 1$, the higher-order contributions in $n$ are increasingly negligible, as seen from the scaling procedure

\begin{equation}\label{scaling} 
h\rightarrow m^{-2}h\,\,\,,\,\,\,\partial_t\rightarrow m^2\partial_t\,\,,
\end{equation}
which leaves $v(t)$ invariant, and attaches a factor $m^{2n-2}$ to the contributions $\propto\partial_t^{\,n-1}v(t)$. In the so-called quasi-stationary approximation all terms of order $n\ge 2$ are neglected. The scenario, developed by Carrard et al. \cite{CGZK}, is based on this procedure, applied to a phenomenological model where a low-velocity dendritic branch is added to the curve $F=F_P(v)$ for a planar interface. Without this additional dendritic branch all trajectories in the $F,v$ -plane would inevitably run to $v=0$. The phase-field simulations by Conti \cite{C1} effectively include all $n$-order terms in Eq. (\ref{expansion}), and for the planar interface lead to the appearance of limit cycles, which are well separated from the line $v=0$. An almost identical behavior arises, if in Eq. (\ref{expansion}) we just include the term $n=2$ , thereby going one step beyond the quasi-stationary approximation.

After expansion of the exponential in Eq. (\ref{expansion}) up to $n=2$, and collection of all terms in Eq. (\ref{concentration}) depending on $t'$, integration over $\tau\equiv t-t'$ yields

\begin{eqnarray}\label{tau-int}
&&-\,ik\int_{0}^{\infty}d\tau\left[1-ik\,\frac{\tau^2}{2}\dot v\right]\exp{[(-k^2+ik\,v)\,\tau]}
\nonumber\\&&=-\,i\,\frac{1}{k-iv}-\,\frac{1}{(k+i\varepsilon)(k-iv)^3}\,\dot v\,\,.
\end{eqnarray} 
The shift $+\,i\varepsilon$ in the denominator of the last term arises from including a term $-\,\varepsilon\,v$ in the preceding exponential, which regularizes the singular point $k=0$ at the upper bound of the integral. 

Next, we take care of the, so far, ignored contribution $ik(\zeta-\zeta')$ in Eq. (\ref{expansion}), and perform the integrations over $k$ separately for the two final contributions in Eq. (\ref{tau-int}). The resulting equations

\begin{eqnarray}\label{k-int}
&&\int_{-\infty}^{+\infty}\frac{dk}{2\pi}\exp{[ik(\zeta-\zeta')]}\frac{-i}{k-iv}=\\&&\Theta(\zeta-\zeta') \exp{[-v(\zeta-\zeta')]}\,\,,\nonumber\\&&\,\nonumber\\&&\dot v\int_{-\infty}^{+\infty}\frac{dk}{2\pi} \exp{[ik(\zeta-\zeta')]}\frac{-1}{(k+i\varepsilon)(k-iv)^3}=\nonumber\\&&\dot v\frac{1}{2}\, \frac{\partial^2}{\partial v^2} \int_{-\infty}^{+\infty}\frac{dk}{2\pi} \frac{\exp{[ik(\zeta-\zeta')]}} {(k+i\varepsilon)(k-iv)}=\nonumber\\&&\dot v\frac{1}{2}\frac{\partial^2}{\partial v^2}\, \frac{1}{v}\, \Bigl\{\Theta(\zeta-\zeta')\exp{[-v(\zeta-\zeta')]}+\Theta(\zeta'-\zeta)\Bigr\}\nonumber
\end{eqnarray}
have, finally, to be multiplied with $U'(\zeta')$ and integrated over $\zeta'$, in order to evaluate the expression (\ref{concentration}) for the solute concentration in the assumed approximation.

In terms of the stationary concentration profile $C_P$, presented in Eqs. (\ref{balance}), the result for $C(z,t)$ reads

\begin{equation}\label{trapping}
C(z,t)=C_P(\zeta;v)+\,\dot v\frac{1}{2}\frac{\partial^2}{\partial v^2}\frac{1}{v}[C_P(\zeta;v)+C_P(\zeta;0)] \,\,.
\end{equation}
Insertion of this into the first equation in Eqs. (\ref{integro}) leads to the closed equation of motion for $Z(t)$ in the form

\begin{eqnarray}\label{differential}
&&\frac{1}{p}\,v=F_P-m^2(Z-Z_P)\\&&+\,G_P(v)-G_P(0)+\dot v\frac{1}{2}\frac{\partial^2}{\partial v^2}\frac{1}{v}[G_P(v)+G_P(0)]\nonumber
\end{eqnarray}
where $G_P(v)$  has been defined in Eqs. (\ref{balance}). For $v=v_P$ the result (\ref{differential}) consistently reduces to the first line in Eqs. (\ref{balance}). Subtracting the latter from Eq. (\ref{differential}), we find for the displacement 

\begin{equation}\label{height}
h(t)\equiv Z(t)-Z_P(t)
\end{equation}
the simpler differential equation

\begin{equation}\label{oscillator}
M(\dot h(t))\,\ddot h(t)+R(\dot h(t))+m^2\,h(t)=0
\end{equation}
where we have introduced the mass and friction functions

\begin{eqnarray}\label{coefficients}
&&M(\dot h)\equiv -\,\frac{1}{2}\,\frac{\partial^2}{\partial v_P^2}\left[\frac{G_P(v_P+\dot h)+G_P(0)}{v_P+\dot h}\right]\,\,,\nonumber\\
&&R(\dot h)\equiv\frac{1}{p}\,\dot h-G_P(v_P+\dot h)+G_P(v_P)\,\,.
\end{eqnarray}
Together with these definitions, Eq. (\ref{oscillator}) represents one of the central results of the present paper. It describes a damped nonlinear oscillator, general properties of which have been discussed in Ref. \cite{Bogoliubov}.  

We mention that in the limit $v(t)=v_P+\dot h(t)\rightarrow 0$ the inertial term shows, after application of the scaling procedure (\ref{scaling}), the behavior

\begin{equation}\label{inertial}
M(\dot h)\,\ddot h\propto \frac{m^2}{v^3}\,\ddot h\,\,.
\end{equation}
The most singular terms in higher-order contributions turn out to carry a pre-factor $(m^2/v^3)^{n-1}$, so that our oscillator equation is only valid for velocities above the cross-over line

\begin{equation}\label{cross-over}
v^3\propto m^2\,\,.
\end{equation}

We, furthermore, observe that, for small $h$, the second definition in Eq. (\ref{coefficients}) implies the behavior

\begin{equation}\label{linear-friction}
R(\dot h)\equiv\left[\frac{1}{p}-G_P'(v_P)\right]\dot h+O(h^2)
\end{equation}
for the friction term in Eq. (\ref{oscillator}), so that, due to the first line in Eqs. (\ref{balance}),

\begin{equation}\label{linear-oscillator}
M(0)\ddot h+F_P'(v_P)\dot h+m^2\,h+O(h^2)=0\,\,,
\end{equation}
in agreement with our linear stability analysis.

The nonlinear differential equation (\ref{oscillator}) obviously has the trivial solution $h(t)=0$, which, however, according to Eq. (\ref{gap}), is unaffected by the Cahn instability in the regime $F_P'(v_P)>0$. In the regime $F_P'(v_P)<0$ we will find solutions $h(t)$, showing an oscillatory behavior in the limit $t\rightarrow\infty$. This behavior is shared by the solute concentration $C(Z(t),t)$ at the oscillating interface, as can be seen from Eq. (\ref{trapping}), taken at $\zeta=0$. 

\section{Limit-Cycle Solutions}

The equations (\ref{trapping}) and (\ref{oscillator}) are valid for a whole class of models with varying equilibrium-concentration profiles $U(z-Z)$. In order to obtain explicit solutions $h(t)$ and $C(Z(t),t)$, we choose the model

\begin{eqnarray}\label{kink}
U(z-Z)&=&\Theta(Z-z)\exp{(z-Z)}\\&+&\Theta(z-Z)[2-\exp{(Z-z)}]\,\,,\nonumber 
\end{eqnarray}
derived in Ref. \cite{PRE} from a two-parabola phase-field model. As also explained in Ref. \cite{PRE}, the choice (\ref{kink}) leads to the expressions

\begin{equation}\label{C-G}
C_P(0,v)=\frac{1}{v+1}\,\,\,,\,\,\,G_P(v)=-\,\gamma\frac{v+2}{(v+1)^2}\,\,.
\end{equation}

\begin{center}
\begin{figure}[h]
 \includegraphics[width=8cm]{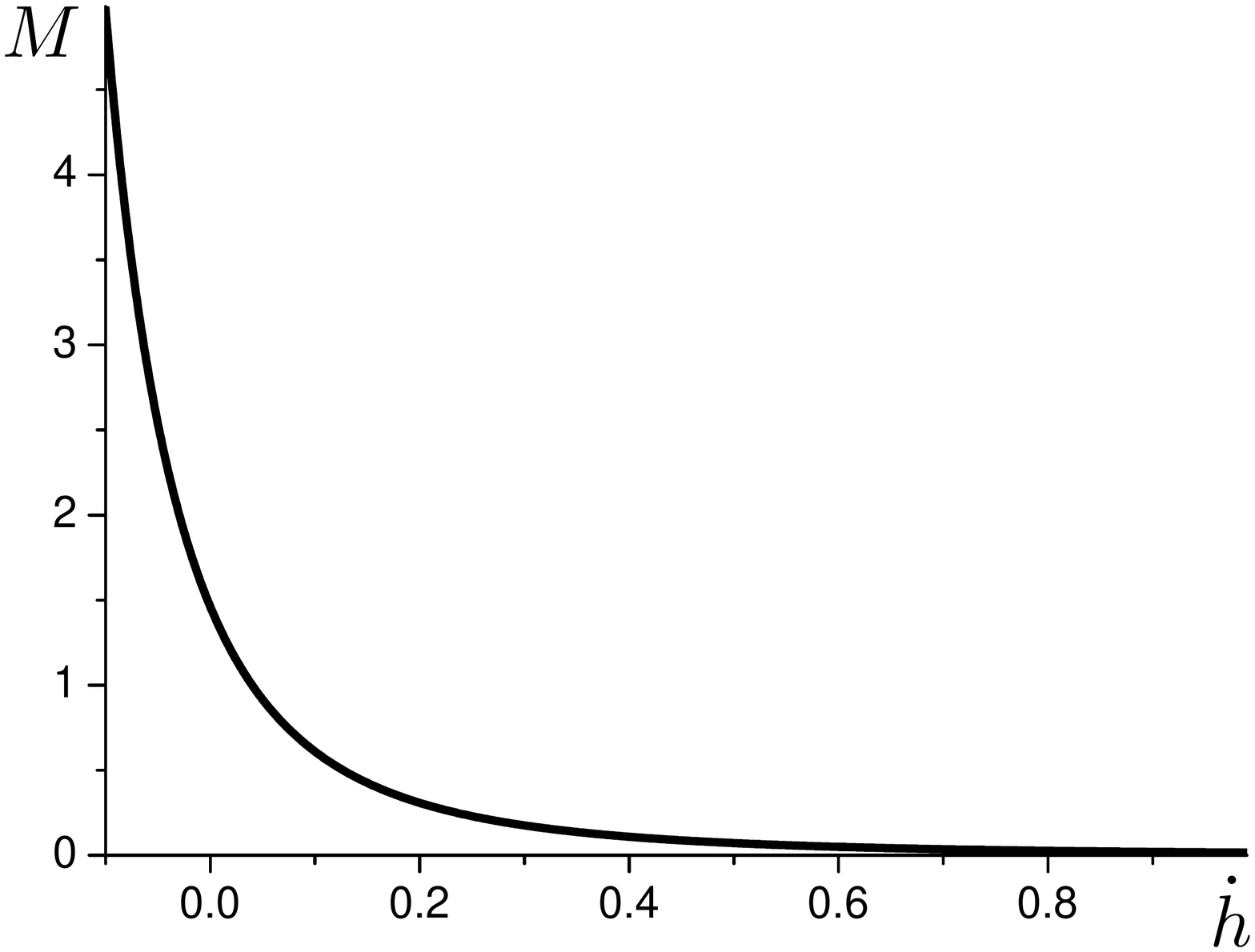}
 \includegraphics[width=8cm]{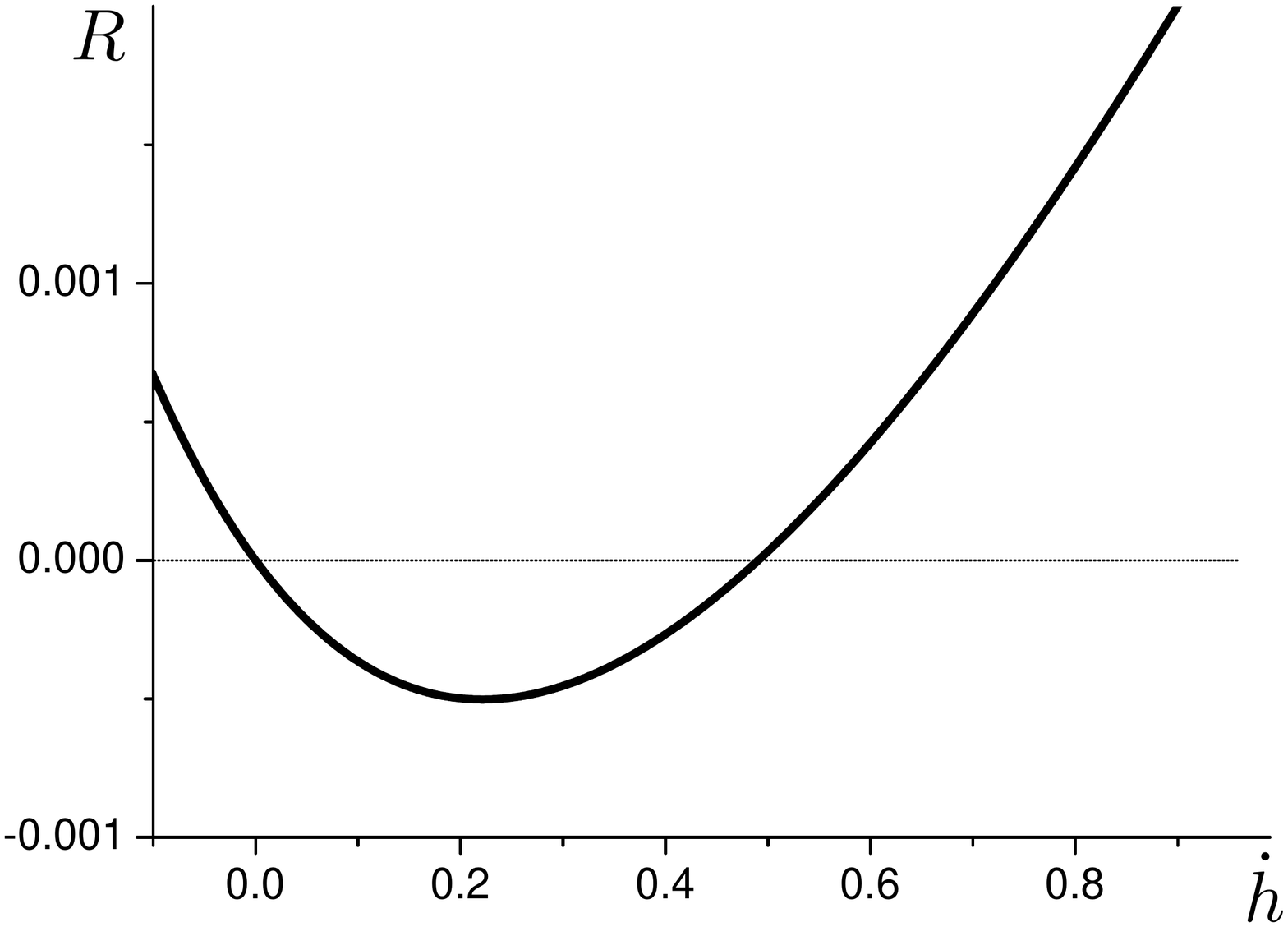}
 \caption{The functions $M(\dot h),R(\dot h)$ for $\gamma=0.01$, $p=100$, and $v_P=0.3$, resulting from the model (\ref{kink}) of the solute concentration.}
\end{figure}
\end{center}

The latter result allows us to determine the quantities (\ref{coefficients}), which e. g. for $\gamma=0.01,p=100,v_P=0.3$ have the form, shown in Fig. 2. A conspicuous property of the function $M(\dot h)$ is its monotonous growth with decreasing velocity, which decisively affects the solutions $h(t)$ in this regime, and, therefore, supports the procedure to include the inertial term in the oscillator equation (\ref{oscillator}). Another implication of Fig. 2 is that, for the present choice of the model parameters, the function $R(\dot h)$ is negative in some finite region where the solution $h(t)=0$ is unstable. The first result in Eqs. (\ref{C-G}), finally, permits to calculate the solute concentration (\ref{trapping}) at the interface, once the solution h(t) of Eq. (\ref{oscillator}) has been found.

Numerically obtained solutions $h(t)$ for the parameter values $\gamma=0.01,p=100,m=0.003$, and for the pulling velocities $v_P=0.522$ and $v_P=0.52$ are shown in Fig. 3. The threshold condition $F_P'(v_C)=0$ generally defines a critical velocity, which in the present case has the value $v_C\approx 0.521$. Above $v_C$ the oscillating trajectories $h(t)$ converge to the value $h(\infty)=0$ whereas below $v_C$ they approach a limit cycle. Fig. 4 shows the same behavior further away from $v_C$, so that, comparing these figures, one observes a kind of critical slowing down.

\begin{center}
\begin{figure}[t]
 \includegraphics[width=8cm]{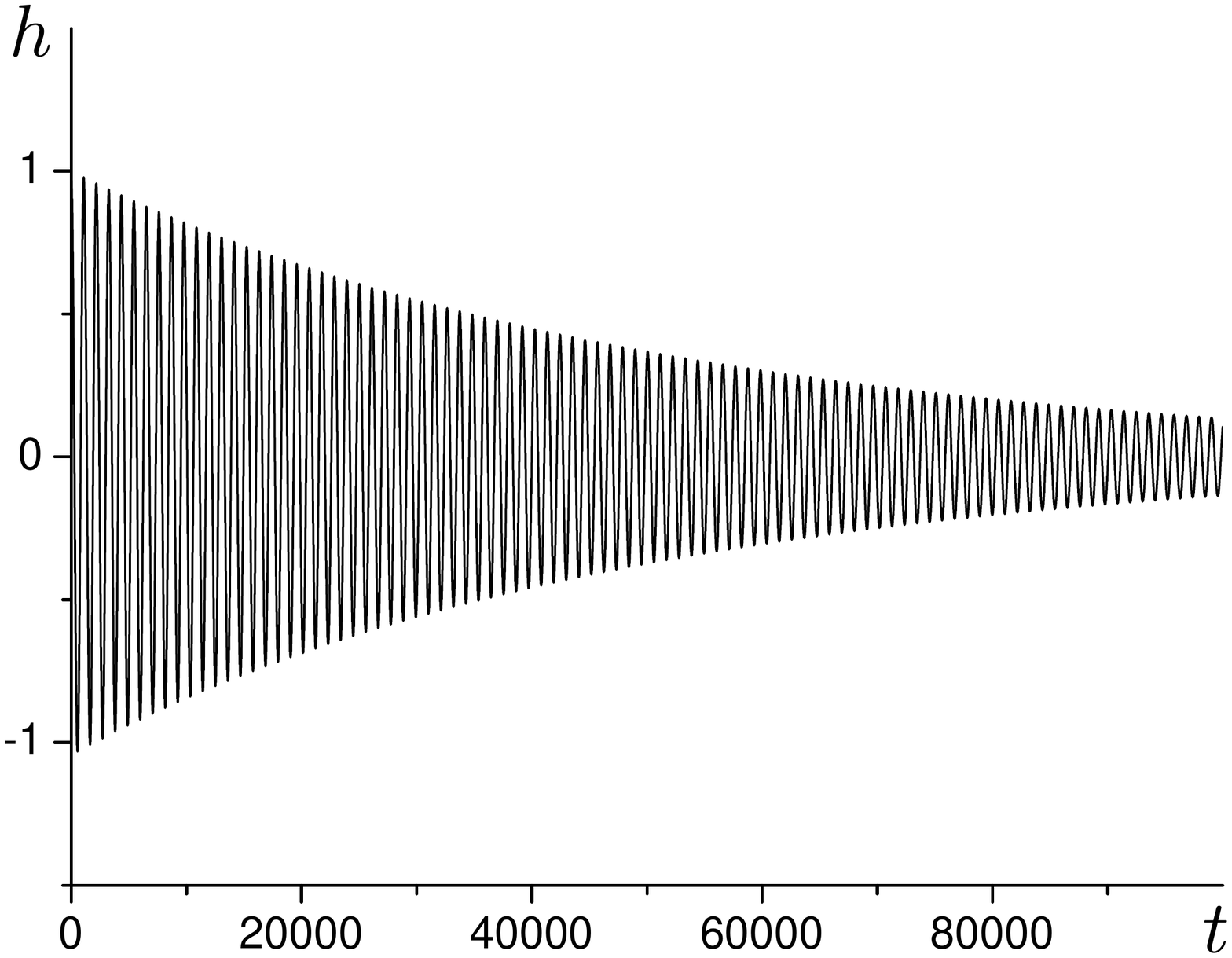}
 \includegraphics[width=8cm]{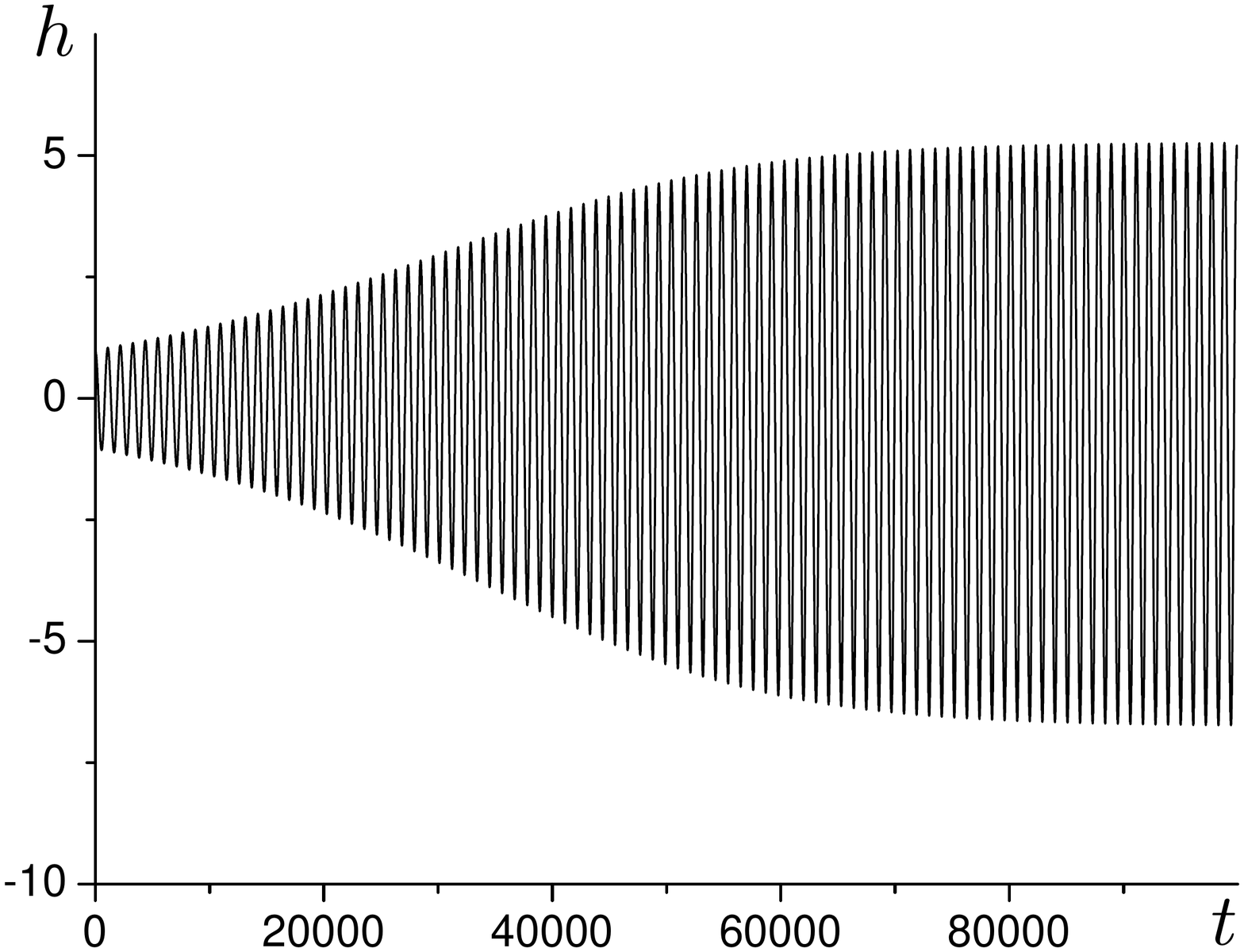}
 \caption{Solutions $h(t)$ for $\gamma=0.01$, $p=100$, $m=0.003$, and pulling velocities $v_P=0.522$, and $v=0.52$.}
\end{figure}
\end{center} 

In the regime $\vert v_P-v_C\vert/v_C\ll 1$ the envelopes in Fig. 3 can be calculated analytically by the Bogoliubov-Mitropolsky procedure \cite{Bogoliubov}, the application of which to the present case is described in Appendix B. The result for the solution of Eq. (\ref{oscillator}) then is found to be

\begin{equation}\label{B-M}
h(t)=a(t)\,cos\,\psi(t)
\end{equation}
where $\psi(t)$ is a rapidly oscillating phase, and $a(t)$ is an amplitude, obeying the differential equation
\begin{equation}\label{amplitude}
\frac{da}{dt}=-\rho_1\,a-\rho_3\,a^3\,\,.
\end{equation}
Here, $\rho_1\equiv r_1(v_P-v_C)$, and the coefficients $r_1,\rho_3$ are determined by the values of the model parameters $\gamma,p,m$. Eq. (\ref{amplitude}) has the solution

\begin{equation}\label{a-solution}
a(t)=a_0\left\{\left[1+\frac{\rho_3}{\rho_1}a_0^2\right]\exp{[2\rho_1\,t]}
-\frac{\rho_3}{\rho_1}a_0^2\right\}^{-1/2}\,\,,
\end{equation}
which for $\rho_1>0$ and $\rho_1<0$ describes the envelopes in Fig. 3. The asymptotic value of the limit-cycle amplitude
 
\begin{center}
\begin{figure}[t]
 \includegraphics[width=8cm]{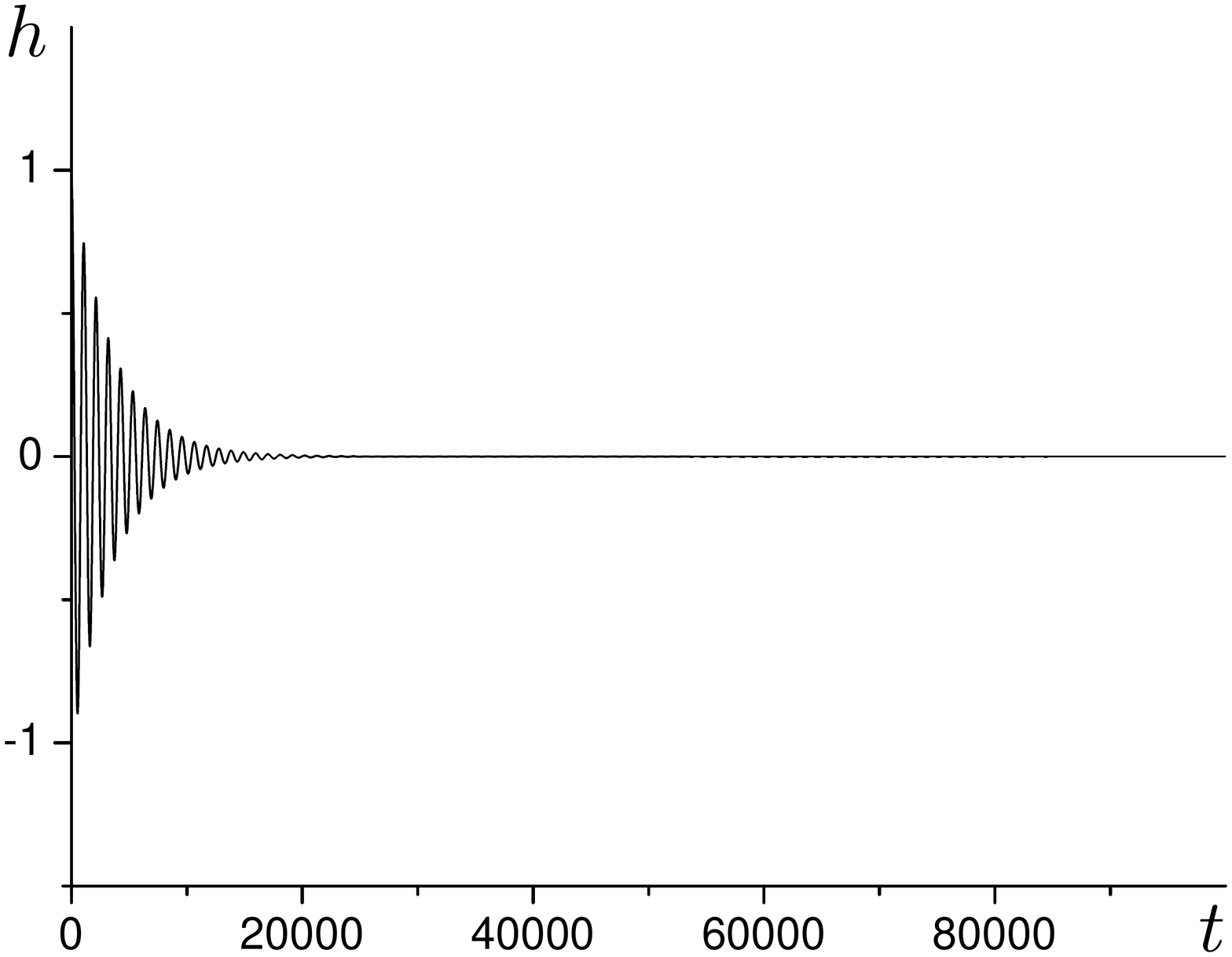}
 \includegraphics[width=8cm]{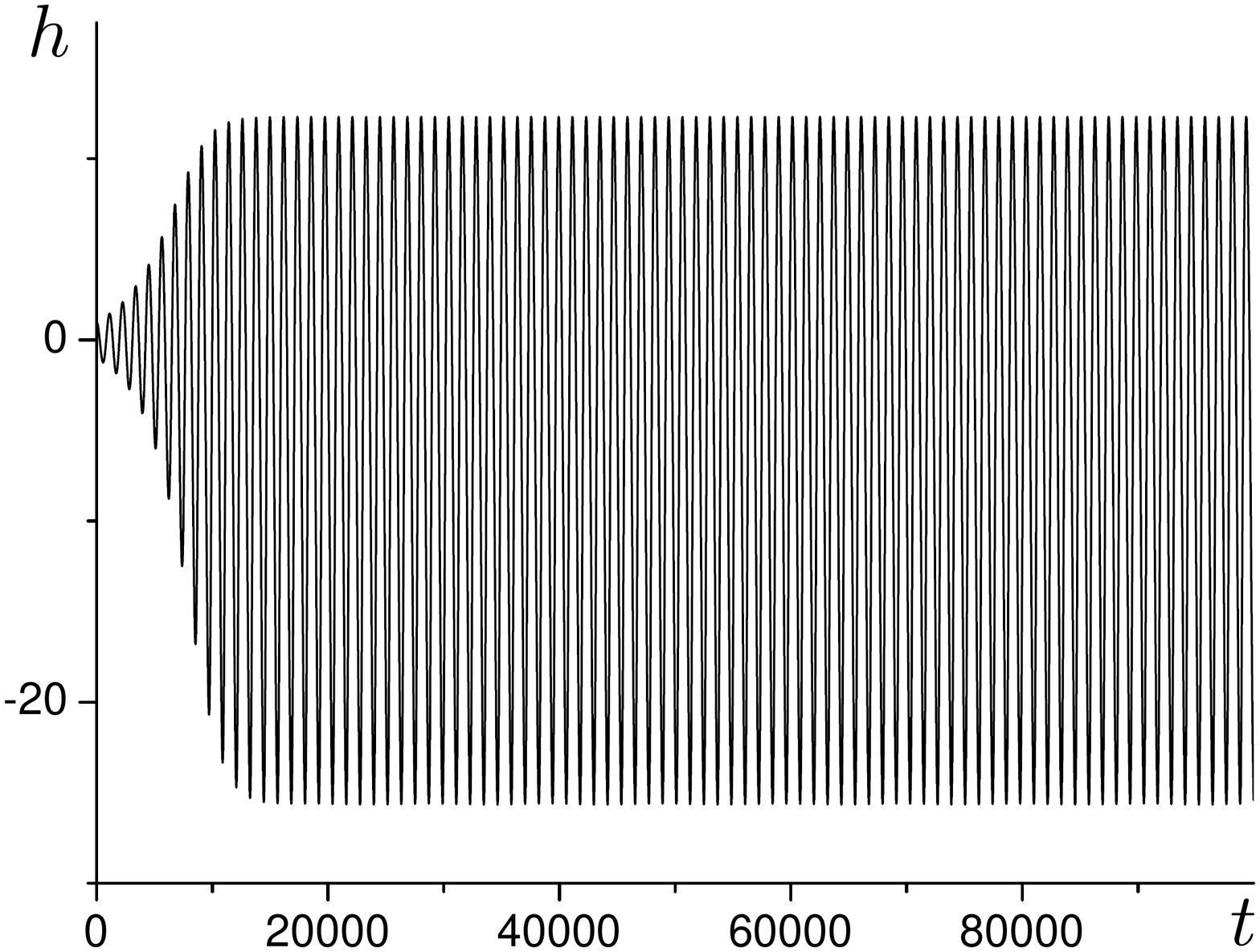}
 \caption{Solutions $h(t)$ for $\gamma=0.01,p=100,m=0.003$, and pulling velocities $v_P=0.53$, and $v_P=0.51$.}
\end{figure}
\end{center}
shows the critical behavior $a(\infty)=\sqrt{-\rho_1/\rho_3}$. In the marginal case $\rho_1=0$ Eq. (\ref{a-solution}) implies the algebraic decay

\begin{equation}\label{marginal}
a(t)=\frac{a_0}{\sqrt{1+2\rho_3(a(0))^2t}}\,\,.
\end{equation}

The rapid oscillations in a fully developed limit cycle, gleaming through in Fig. 4, are most suitably analyzed by numerical computations. A first result is the orbit of a limit cycle in the $h,\dot h$ - plane, shown in Fig. 5 for the parameter values $\gamma=0.01,p=100,m=0.003$, and the pulling velocity $v_P=0.5$. The related oscillations of the trajectories $h(t),\dot h(t)$, and of $C(Z(t),t)$ are displayed in Fig. 6. Since the term $m^2h(t)$ measures the temperature at the oscillating interface, this quantity is effectively also included in Fig. 6. 

By reducing the pulling velocity to the value $v_P=0.3$ at constant parameters $\gamma,p,m$, one obtains the shape of the trajectories $h(t),\dot h(t),C(Z(t),t)$ deeper inside the limit-cycle regime. The results for $h(t),\dot h(t)$, displayed in Fig. 7, are remarkably close to the findings by Conti 

\begin{center}
\begin{figure}[h]
 \includegraphics[width=8cm]{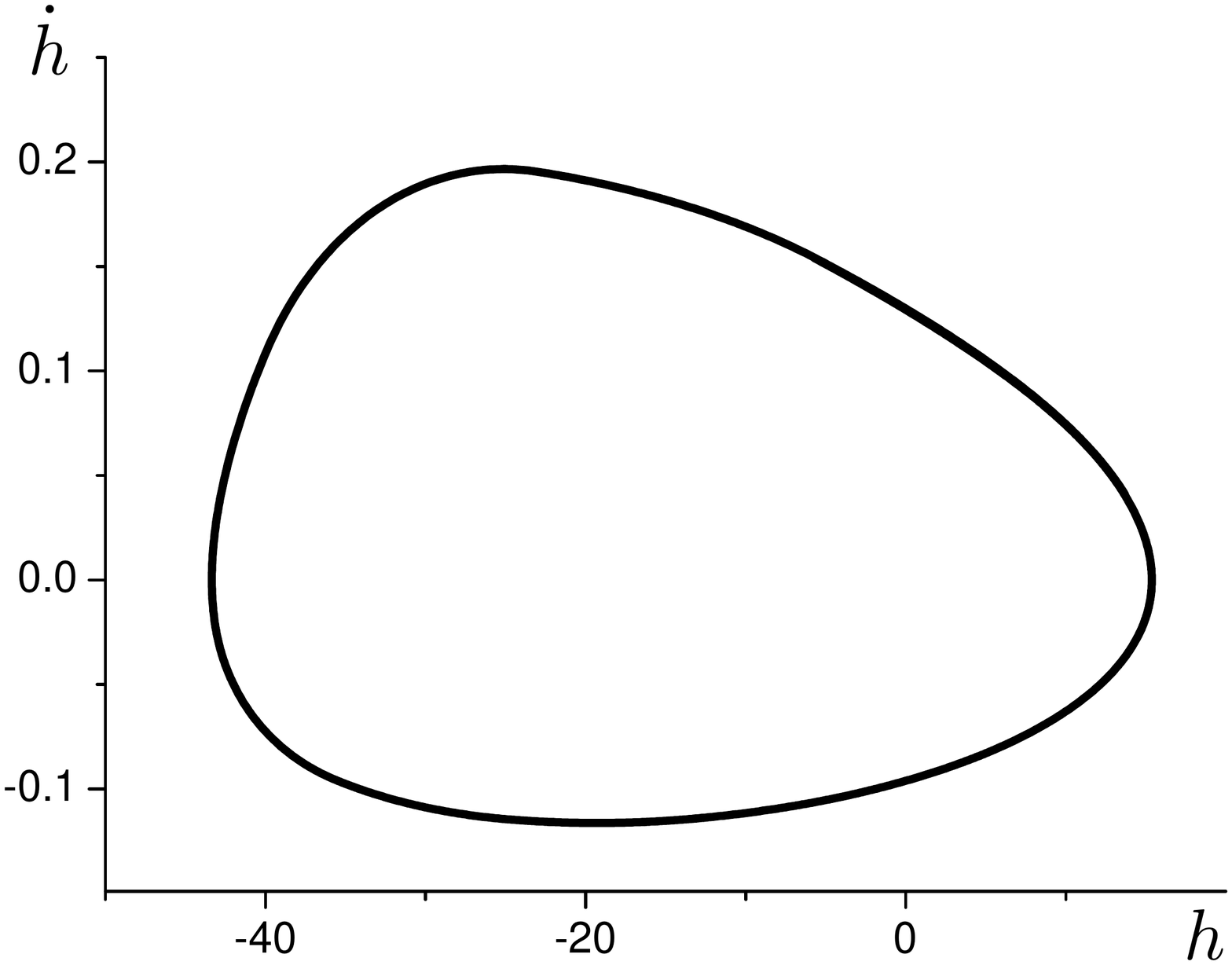}
 \caption{Orbit of the cycle for $\gamma=0.01,p=100,m=0.003$, and $v_P=0.5$.}
\end{figure}
\end{center}
in Ref. \cite{C1}. From the associated behavior of $C(Z(t),t)$ we, moreover, see that the transitions between high- and low-concentration layers are joined by large-acceleration sections. As already pointed out by Carrad et al. \cite{CGZK}, this explains the appearance of relatively sharp interfaces between these layers. Fig. 8, finally, presents our result for the orbit of the limit cycle belonging to Fig. 7.

\section{Banded-Structure Formation}

The layer formation, induced by the above limit-cycle solutions is unaffected by the Mullins-Sekerka instability, deriving from Eq. (\ref{dispersion}). This follows from Fig. 9, which shows the neutral stability lines, enclosing the unstable regions of the Cahn and the Mullins-Sekerka instabilities, and the projection of the limit cycle in Fig. 8. We have to point out, however, that the form of the the Mullins-Sekerka neutral line is only an approximate one, since it is related to a steady-state reference motion with velocity $v_P$. The approximation seems, however, to be acceptable due to the almost stationary behavior of $\dot h(t)$ in Fig. 7 at low velocities. We, accordingly, expect that Fig. 7 induces the formation of precipitation-free solute bands.

The approximation for the Mullins-Sekerka neutral line is apparently more justified for the limit cycle, belonging to Fig. 10. In this case, the projection of the cycle enters the unstable region of the Mulins-Sekerka instability, as seen in Fig. 11. Accordingly, the interface will develop a dendritic microstructure at low velocities, which dissolves again in the high-velocity regime. This is just what one expects to happen in the creation of banded structures, and it is in agreement with the simulations in Ref. \cite{C2}.

\begin{center}
\begin{figure}[t]
 \includegraphics[width=8cm]{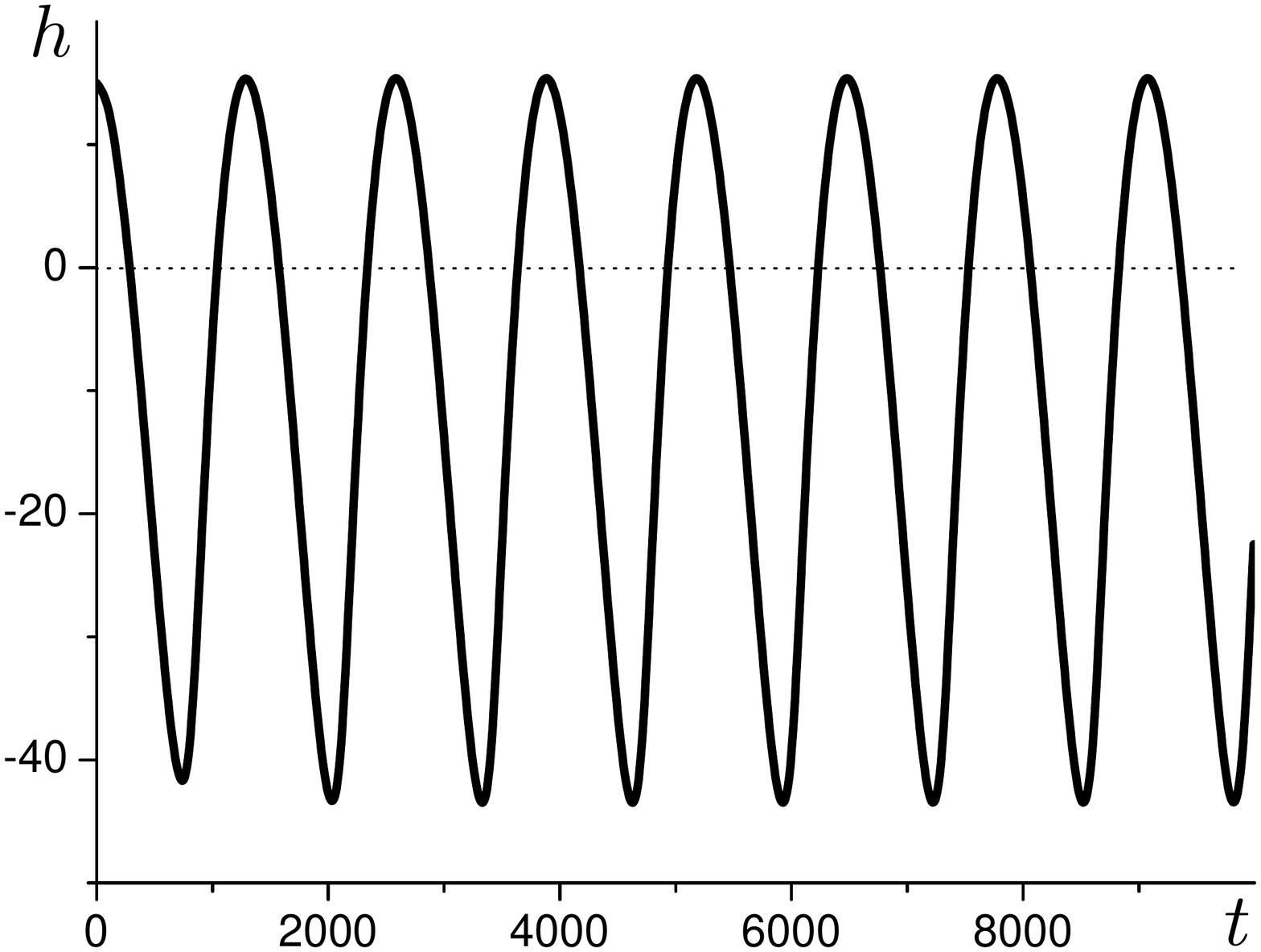}
 \includegraphics[width=8cm]{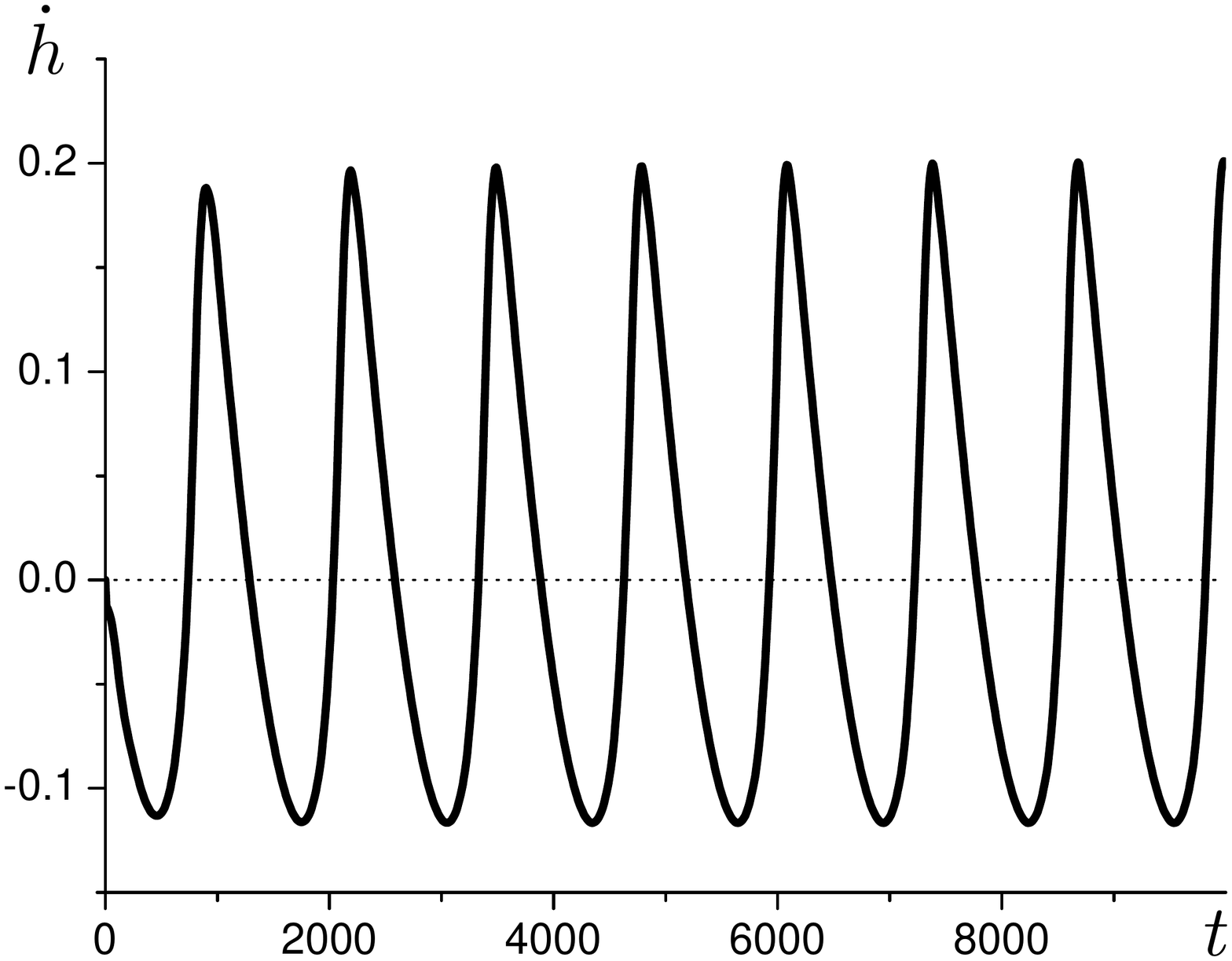}
 \includegraphics[width=8cm]{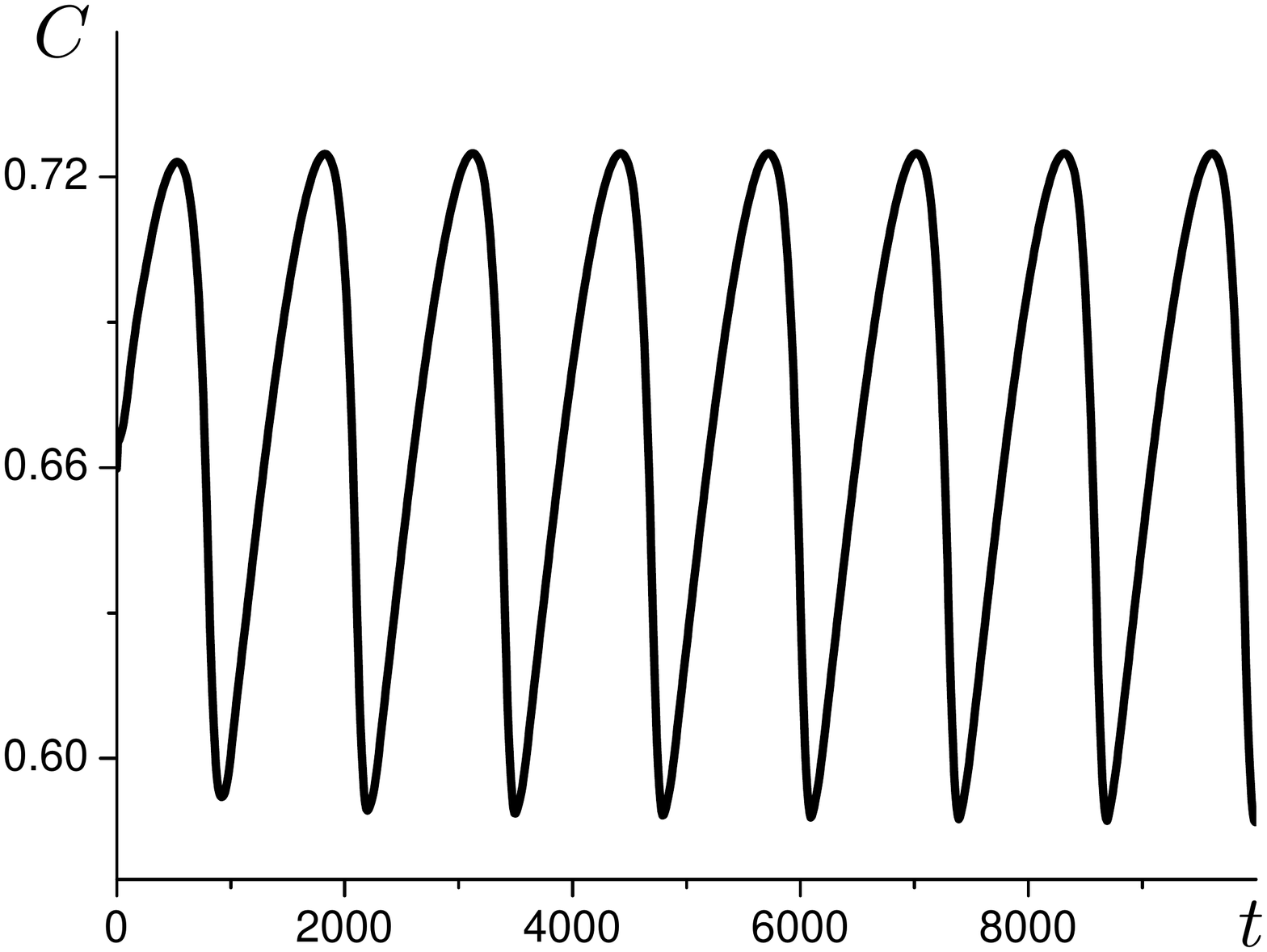}
 \caption{Solutions $h(t),\dot h(t),C(Z(t),t)$ for $\gamma=0.01$, $p=100$, $m=0.003$, and $v_P=0.5$.}
\end{figure}
\end{center}

\begin{center}
\begin{figure}[t]
 \includegraphics[width=8cm]{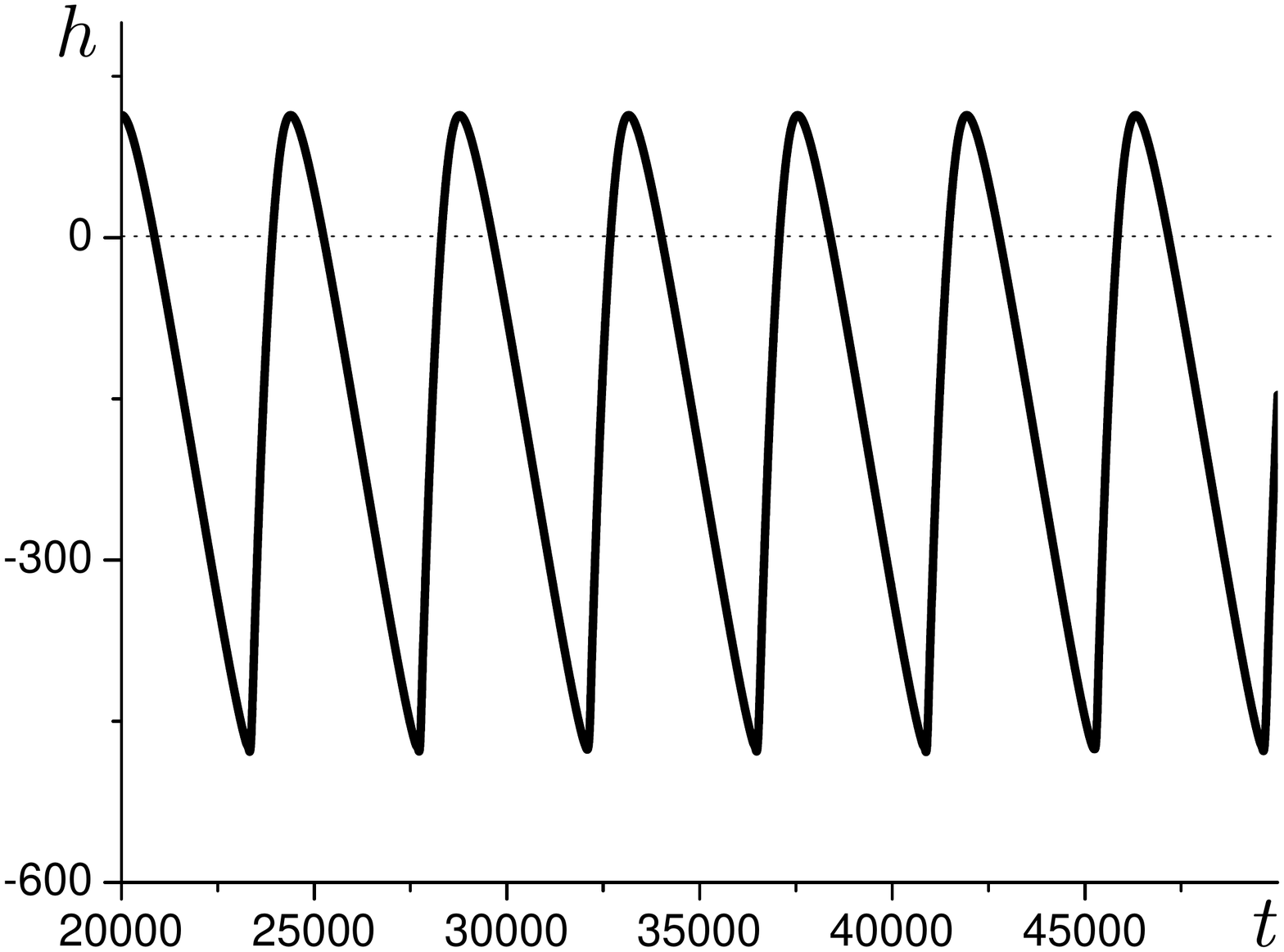}
 \includegraphics[width=8cm]{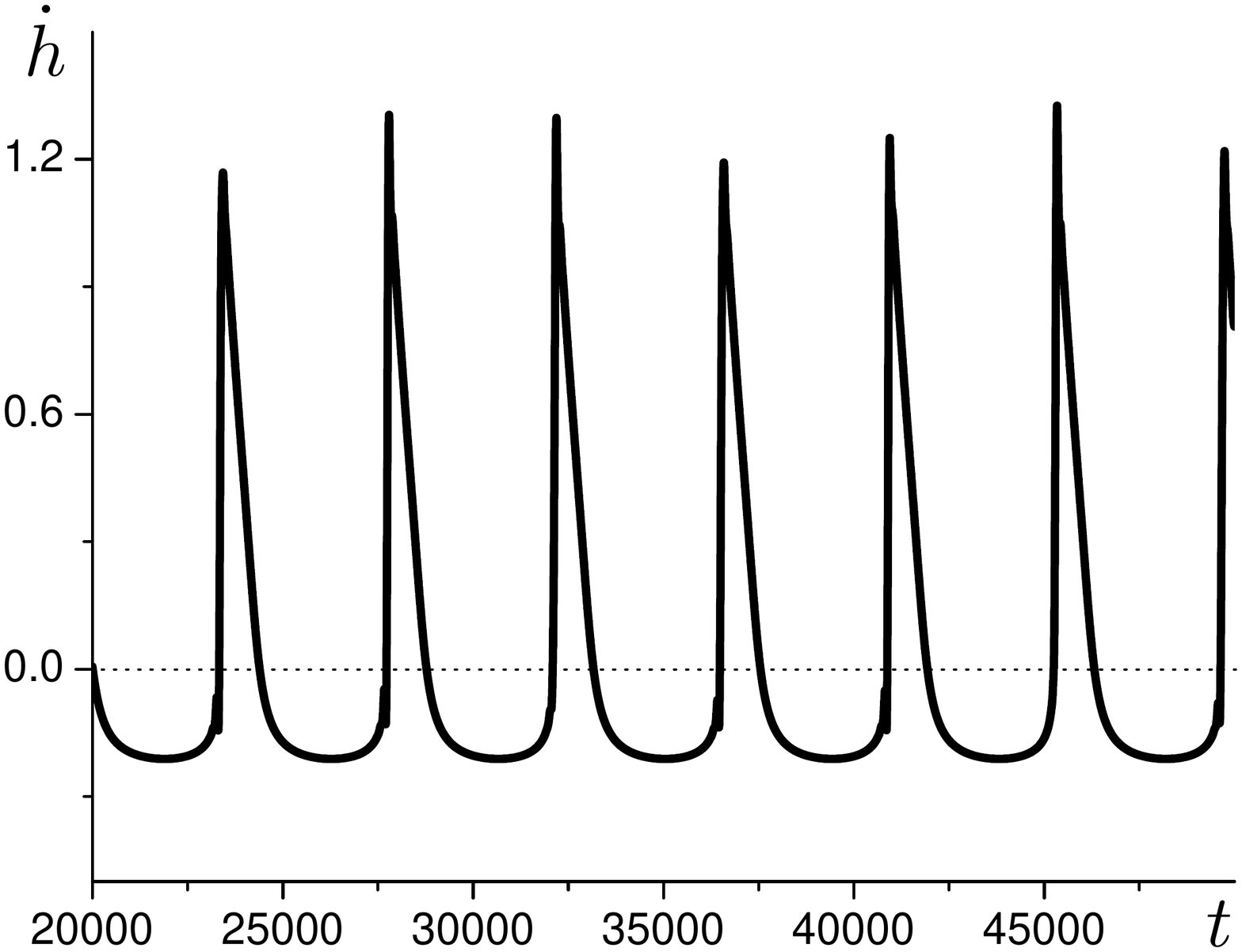}
 \includegraphics[width=8cm]{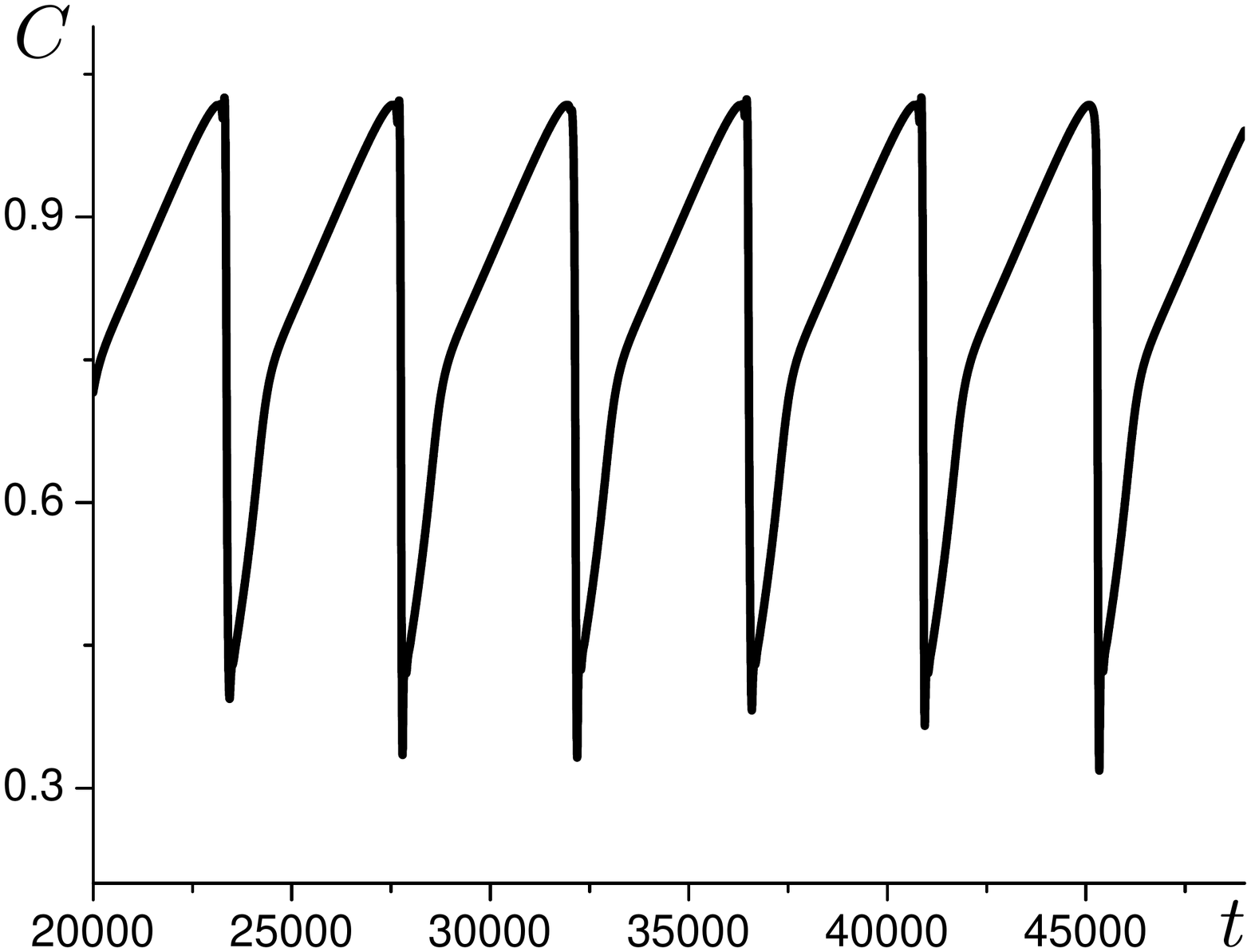}
 \caption{Solutions $h(t),\dot h(t),C(Z(t),t)$ for $\gamma=0.01$, $p=100$, $m=0.003$, and $v_P=0.3$.}
\end{figure}
\end{center}

\begin{center}
\begin{figure}[t]
 \includegraphics[width=8cm]{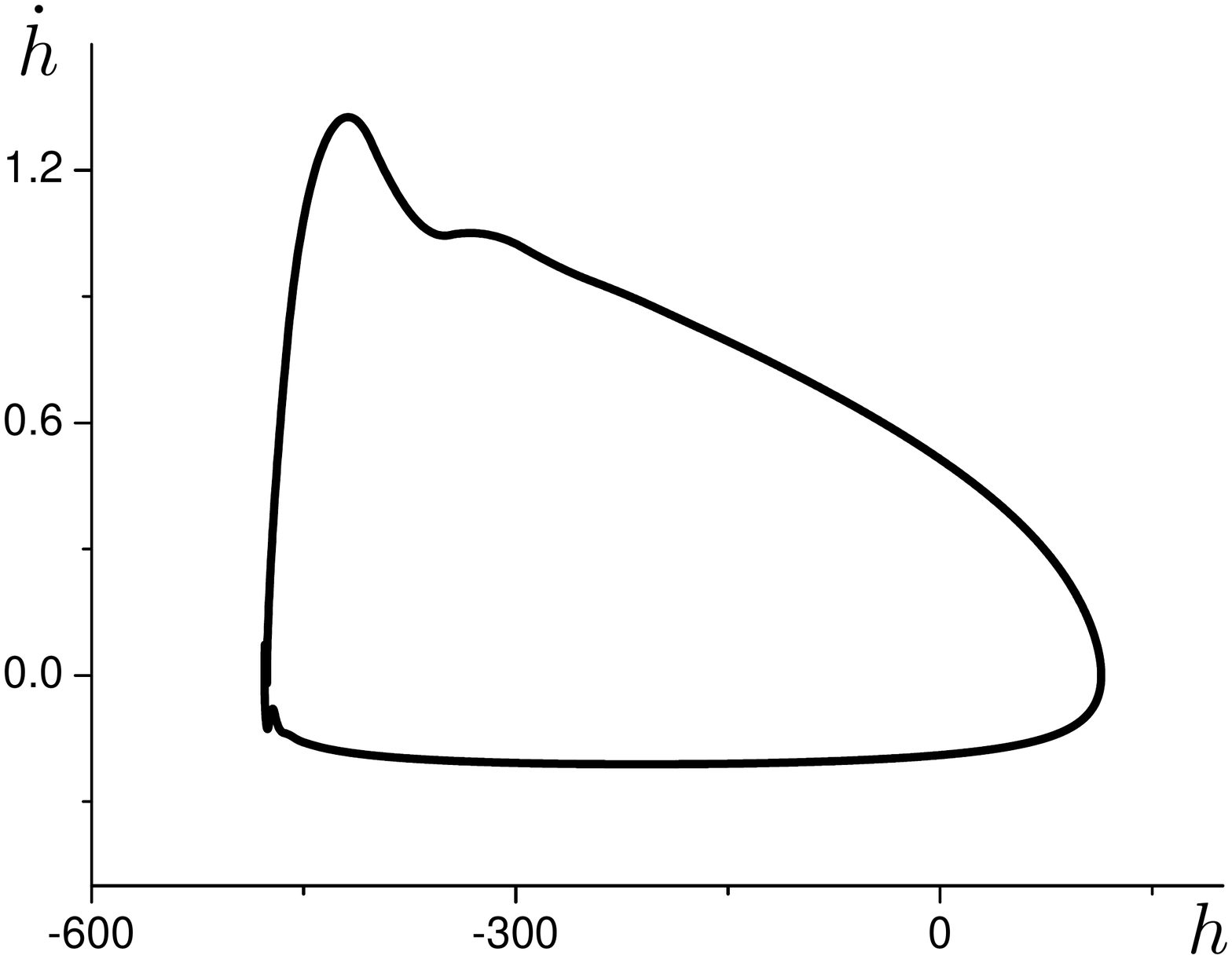}
 \caption{Orbit of the cycle for $\gamma=0.01,p=100,m=0.003$, and $v_P=0.3$.}
\end{figure}
\end{center}

\begin{center}
\begin{figure}[t]
 \includegraphics[width=8cm]{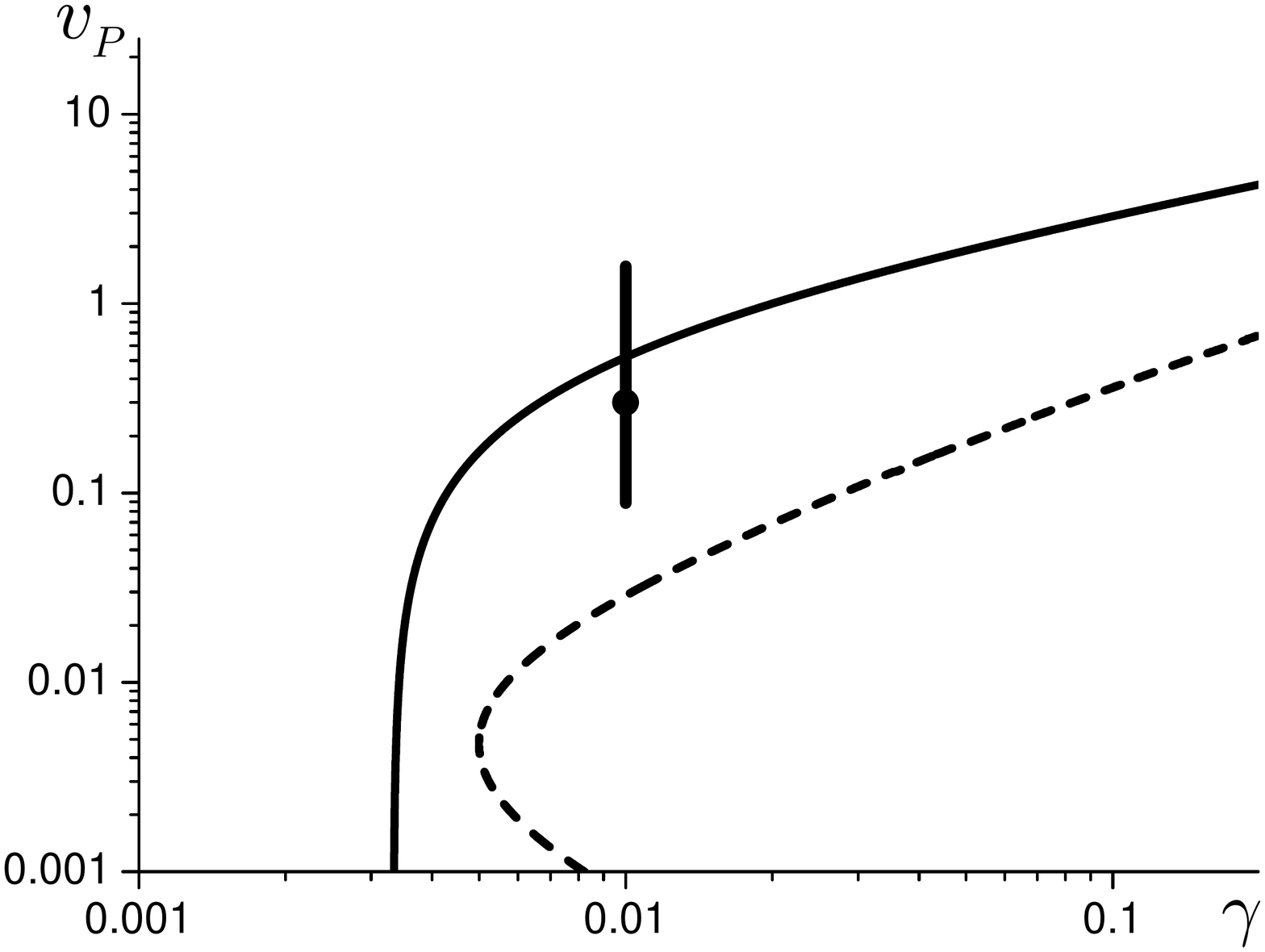}
 \caption{Neutral lines, enclosing the regions of the Cahn (solid line), and of the Mullins-Sekerka (dashed line) instability. The vertical line is the projection of the limit cycles in Fig. 9.}
\end{figure}
\end{center}

\begin{center}
\begin{figure}[h]
 \includegraphics[width=8cm]{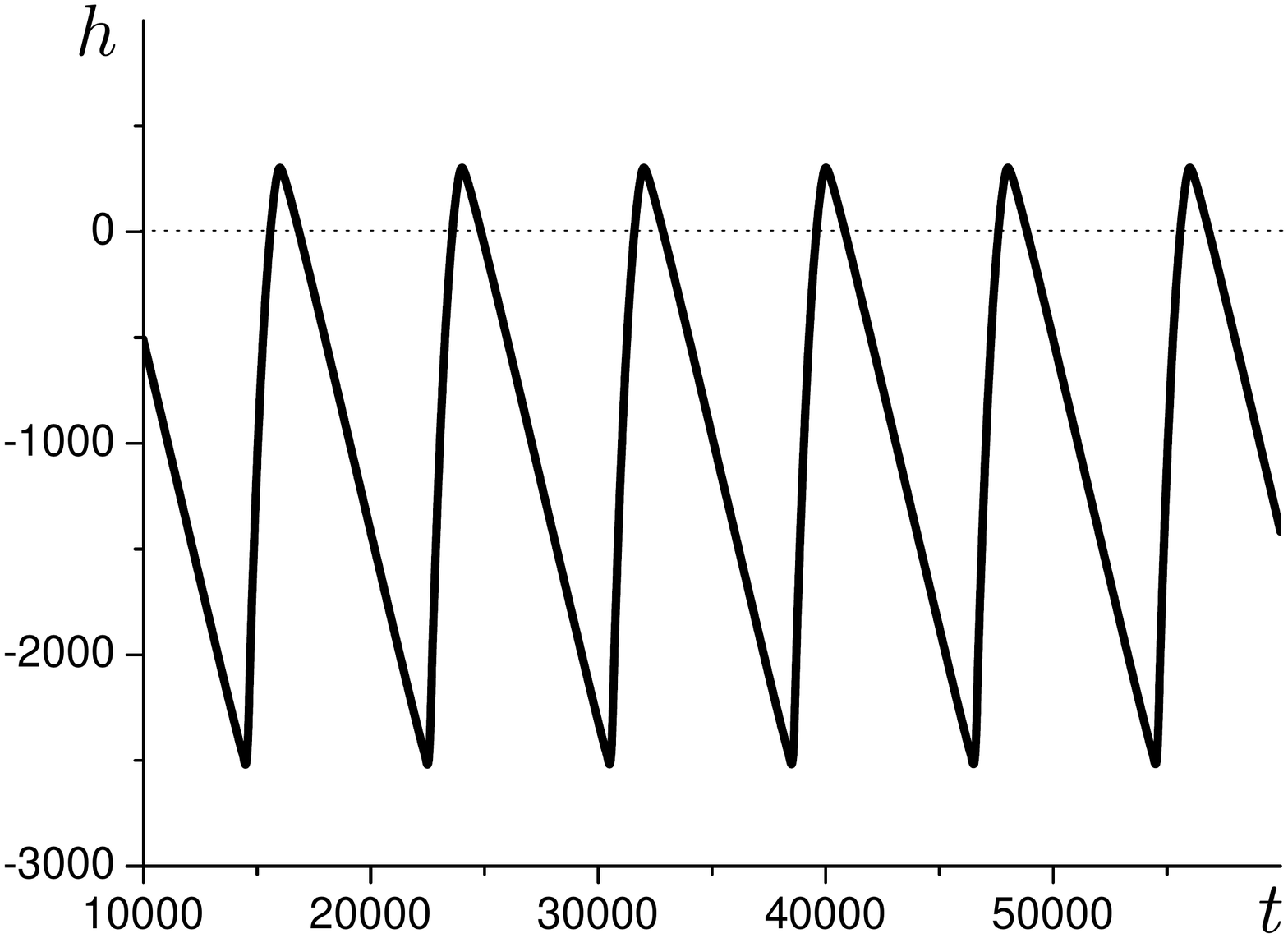}
 \includegraphics[width=8cm]{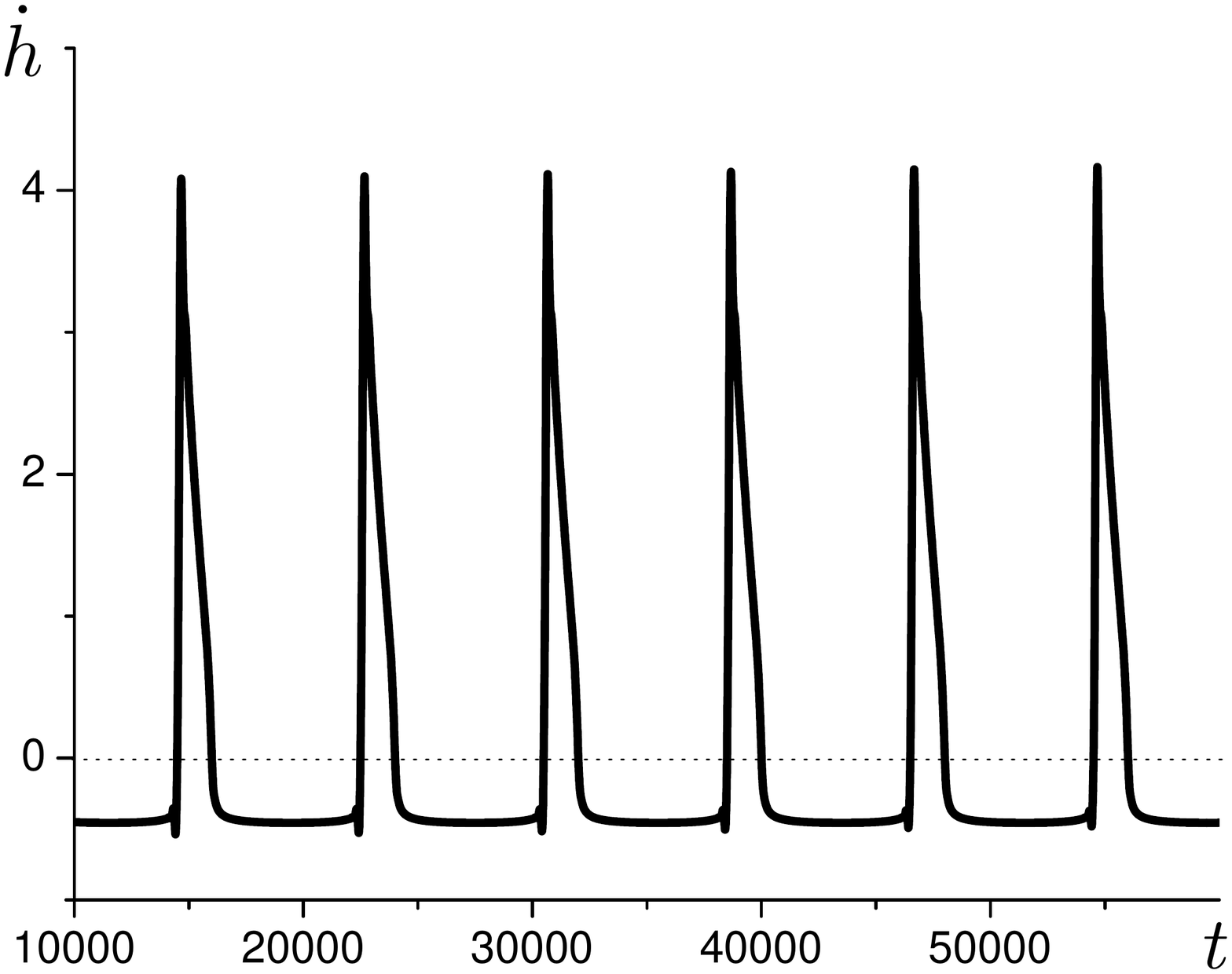}
 \includegraphics[width=8cm]{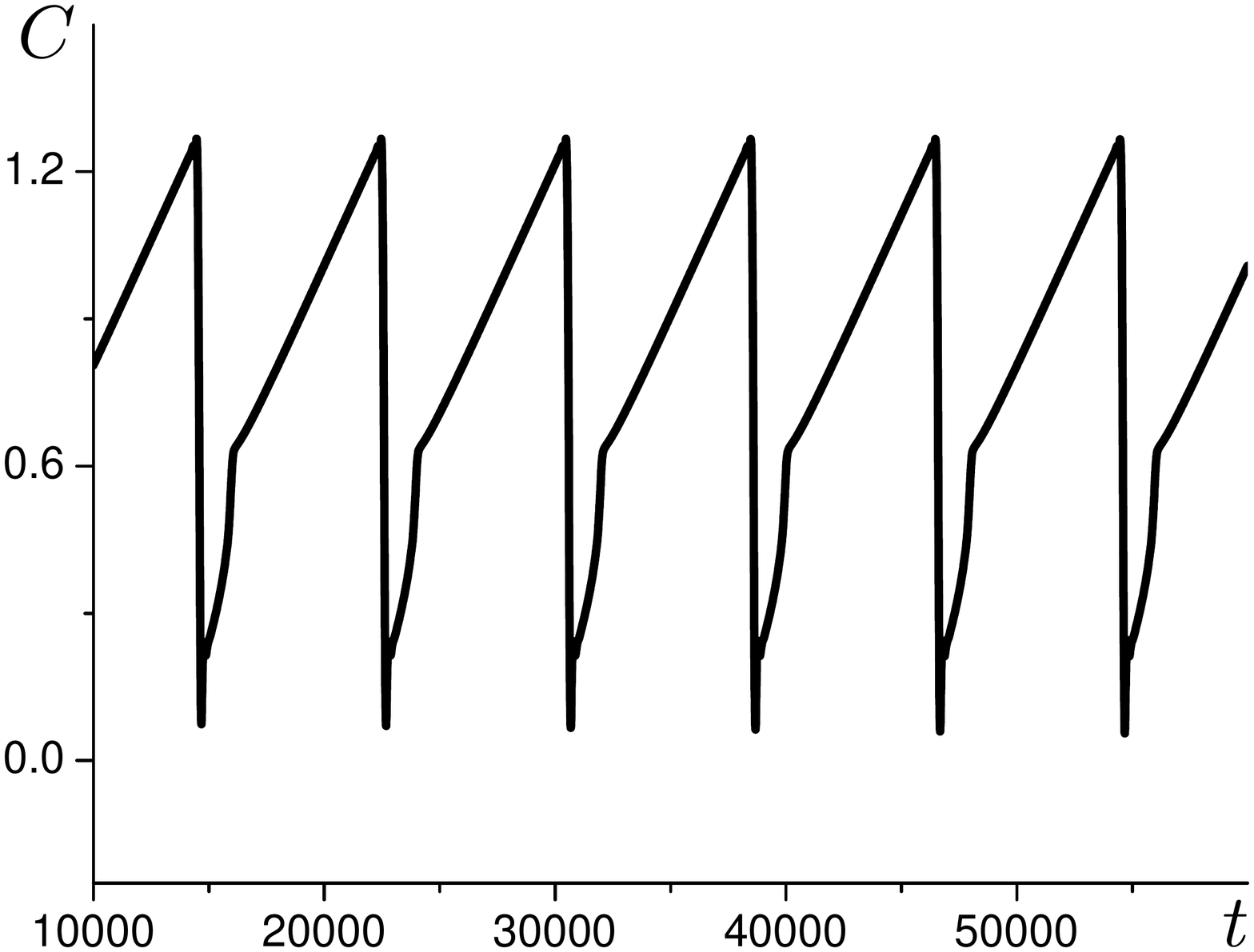}
 \caption{Solutions $h(t),\dot h(t),C(Z(t),t)$ for $\gamma=0.02$, $p=100$, $m=0.003$, and $v_P=0.5$.}
\end{figure}
\end{center}

\begin{center}
\begin{figure}[h]
 \includegraphics[width=8cm]{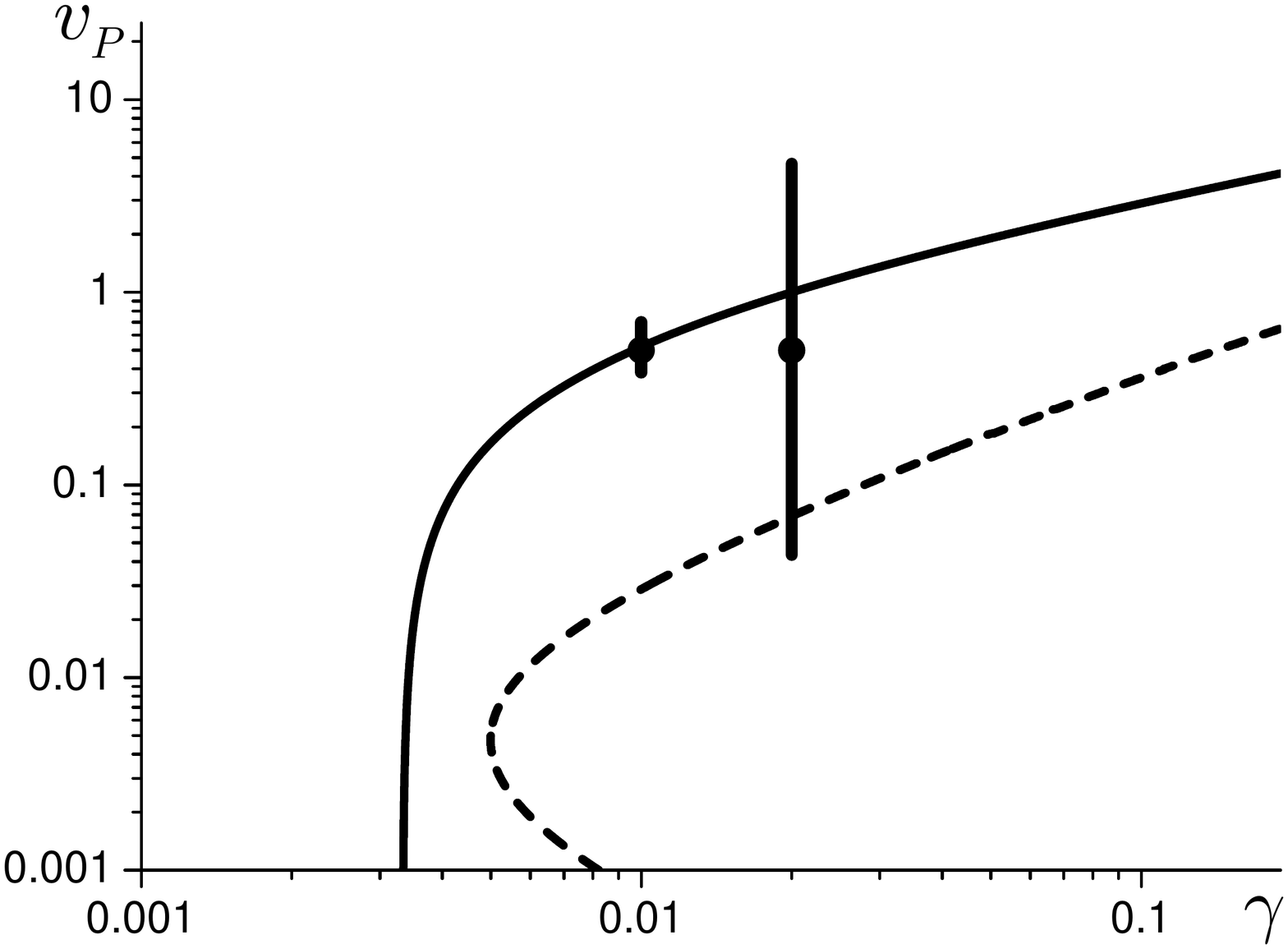}
 \caption{Neutral lines of the Cahn (solid line), and of the Mullins-Sekerka (dashed line) instability. The vertical lines are the projections of the limit cycles in Figs. 6 and 10.}
\end{figure}
\end{center}

\section{Discussion}

A crucial point of our analysis is the observation that the periodic motion of a planar solidification can only be explained, if we go one step beyond the quasi-stationary approximation in the expansion (\ref{expansion}). For a quantitative evaluation higher-order terms can be neglected, because the definition of the effective expansion parameter $m^2$ in Eqs. (\ref{dimensionless}) implies $m^2\approx 5\times 10^{-5}$, if we adopt from Ref. \cite{CGZK} the value $S=2\times 10^5K/cm$ of the temperature gradient, and from Ref. \cite{C1} the material parameters $T_M=1728\,K, L=2350\,J/cm^3, \sigma=3.7\times 10^{-5}J/cm^2$ for Nickel, and the interface thickness $2\xi=1.68\times 10^{-7}cm$.

The fact that the restoring force in Eq. (\ref{oscillator}) is given by $m^2 h(t)$ may raise the suspicion that the presence of a temperature gradient is an essential ingredient of our theory. This is only true, however, for a planar geometry of the solidification front. In case of a growing spherical nucleus the parameter $m^2$ turns out to be proportional to the ratio $\xi/R_c$ where $R_c$ is the critical radius of the droplet.

Generally, the parameter $m$ determines the period $\Omega_0$ of a limit cycle. Close to the threshold at $v_C$ the friction term in Eq. (\ref{oscillator}) can be neglected, so that, due to scaling, $\Omega_0\propto m$. Deep inside the limit-cycle regime accelerations are negligible in most parts of the trajectories $h(t)$ in Figs. 7 and 10, suggesting to neglect the inertial term in Eq. (\ref{oscillator}). Its scaling behavior then implies $\Omega_0\propto m^2$, in accordance with the statement in Ref. \cite{CGZK} that the band width in a pronounced banded structure is inversely proportional to the temperature gradient.

As a final point we note that most phenomenological approaches are based on the assumption of an $N$-shaped force-velocity relation. This suggests the formation of a hysteresis loop, which is considered to represent the limit cycle, describing the defect oscillations. In our model the driving force is a convex function of velocity, excluding the existence of a hysteresis loop. Instead the necessary turnaround of a trajectory at low velocities is provided by the inertial term in the oscillator equation, which proves the importance of including this term in the equation for the interface position.

\section{Appendix A}

Within our capillary-wave approach the most general effective Hamiltonian for the directional solidification of a dilute binary alloy reads

\begin{eqnarray}\label{Hamiltonian-E}
\mathcal{H}&=&\int d^2x\,\frac{\sigma}{2}(\partial Z)^2+\int d^3r\left\{\frac{\kappa_0}{2} \Bigl[C-U(z-Z)\Bigr]^2\right.\nonumber\\&+&\nu\Bigl[C-U(z-Z)\Bigr]\Bigl[\widetilde C-\widetilde U(z-Z)\Bigr]\nonumber\\&+&\left.\frac{\widetilde\kappa}{2}\Bigl[\widetilde C-\widetilde U(z-Z)\Bigr]^2\right\}\,\,.
\end{eqnarray}
Here, we have introduced a field $\widetilde C({\bf r},t)$, which is related to the energy density $E({\bf r},t)$ by the equation

\begin{equation}\label{C-tilde-E}
\widetilde C({\bf r},t)\equiv\left(1-\frac{\nu}{\widetilde\kappa}\,\frac{T_M\Delta C}{L}\right)\frac{E({\bf r},t)}{T_M}\,\,.
\end{equation}
From this and the equilibrium condition $\delta H/\delta\widetilde C=0$ we conclude that $\widetilde U(z-Z)$ obeys the relation

\begin{equation}\label{U-tilde-L}
\widetilde U(+\infty)-\widetilde U(-\infty)=\left(1-\frac{\nu}{\widetilde\kappa}\,\frac{T_M\Delta C}{L}\right)\frac{L}{T_M}\,\,,
\end{equation}
which, remembering the relations (\ref{misc-gap}) and (\ref{C_E-U}), suggests to refine our model by assuming

\begin{equation}\label{U-tilde-U}
\widetilde U(z-Z)=\left(\frac{L}{T_M\Delta C}-\frac{\nu}{\widetilde\kappa}\right)\,\,U(z-Z)\,\,.
\end{equation}
For the derivation of the model (\ref{I-Hamiltonian}) - (\ref{int-temp}) the physical meanings of the coupling constants $\nu,\widetilde\kappa$ are irrelevant, because they will be absorbed into renormalizations of the parameters $\kappa_0$ and $D_0$. The only generally important constraint on the coupling constants is

\begin{equation}\label{positivity}
\kappa_0\widetilde\kappa-\nu^2\ge 0\,\,,
\end{equation}
which ensures stability of the Hamiltonian (\ref{Hamiltonian-E}).

The equations of motion of the generalized model read 

\begin{eqnarray}\label{dynamics-E}
\partial_t Z&=&-\,\Lambda\,\frac{\delta\mathcal{H}}{\delta Z}\,\,,\\\partial_t C&=&D_0\,\nabla^2\frac{1}{\kappa_0}\, \frac{\delta\mathcal{H}}{\delta C}\,\,,\nonumber\\\partial_t\widetilde C&=&\widetilde D\,\,\nabla^2\frac{1}{\,\widetilde\kappa\,\,}\,\frac{\delta\mathcal{H}}{\delta\widetilde C}\,\,,\nonumber
\end{eqnarray}
where $\widetilde D$ is the heat diffusion constant. The relation

\begin{equation}\label{Legendre}
T({\bf r},t)\equiv T_S+\frac{\delta\mathcal{H}}{\delta\widetilde C}
\end{equation}
defines a temperature field via a shifted local Legendre transform of $\widetilde C({\bf r},t)$, obeying the condition $T({\bf r},t)=T_S$ in thermal equilibrium $\delta H/\delta\widetilde C=0$.

In the limiting case of an infinite heat conductivity, $\widetilde D\rightarrow\infty$, and the last of the Eqs. (\ref{dynamics-E}) is solved, for the boundary conditions $T(Z_P)=T_P,\,T'(Z_P)=S$, by the static temperature field

\begin{equation}\label{ext-T}
T(z)=T_P+S(z-Z_P)\,\,.
\end{equation}
Insertion of the Hamiltonian (\ref{Hamiltonian-E}) and the result (\ref{ext-T}) into Eq. (\ref{Legendre}) leads to the relation

\begin{eqnarray}\label{gradient}
&&\widetilde\kappa\Bigl[\widetilde C({\bf r},t)-\widetilde U(z-Z)\Bigr]+\nu\Bigl[C({\bf r},t)-U(z-Z)\Bigr]
\nonumber\\&&=-\Bigl[T_S-T_P+S(z-Z_P)\Bigr]\,\,.
\end{eqnarray}

If the expression for $\widetilde C-\widetilde U$, extracted from Eq. (\ref{gradient}), is inserted into the second of the Eqs. (\ref{dynamics-E}), one recovers the corresponding equation in Eqs. (\ref{I-dynamics}) with the reduced diffusion constant

\begin{equation}\label{diff-const}
D\equiv\left(1-\frac{\nu^2}{\kappa_0\widetilde\kappa}\right)D_0\,\,.
\end{equation}
In the calculation of the force $-\,\delta\mathcal{H}/\delta Z$, entering the first of the Eqs. (\ref{dynamics-E}), the relations (\ref{U-tilde-U}) and (\ref{gradient}) can be used to eliminate the quantities $\widetilde U, \widetilde C$, which leads to the result 

\begin{eqnarray}\label{Z-eqn}
&&-\,\frac{\delta\mathcal{H}}{\delta Z}=\sigma\,\partial^2Z\\
&&-\kappa\int_{-\infty}^{+\infty}dz\,U'(z-Z)\Bigl[C-U(z-Z)\Bigr] \nonumber\\&&+\,\frac{L}{T_M\Delta C}\int_{-\infty}^{+\infty}dz\,U'(z-Z)\Bigl[T_S-T_P+S(z-Z_P)\Bigr]\nonumber
\end{eqnarray}
with the renormalized coupling constant

\begin{equation}\label{coupl-const}
\kappa\equiv\kappa_0-\frac{\nu^2}{\widetilde\kappa}\,\,.
\end{equation}
The first two terms on the right-hand side of Eq. (\ref{Z-eqn}) are identical to the force $-\,\delta H/\delta Z$ in the first equation in Eqs. (\ref{I-dynamics}). Assuming that $U'(\zeta)$ is an even function, as in case of the model (\ref{kink}), the last integral in Eq. (\ref{Z-eqn}) reduces to the driving force (\ref{force}). We mention that the coupling term $\propto\nu$ in the Hamiltonian (\ref{I-Hamiltonian}) only gives rise to a shift in Eqs. (\ref{C-tilde-E}), (\ref{U-tilde-L}), (\ref{U-tilde-U}), and to parameter renormalizations in Eqs. (\ref{diff-const}), (\ref{coupl-const}), which all disappear in the commonly considered case $\nu=0$.

\section{Appendix B}

In order to derive the differential equation (\ref{amplitude}) for the amplitude $a(t)$, we start from the oscillator equation (\ref{oscillator}), rewritten in the form

\begin{equation}\label{oscill}
\ddot h+\Omega^2h=-\,\frac{R(\dot h)}{M(\dot h)}+m^2\left[\frac{1}{M(0)}-\frac{1}{M(\dot h)}\right]h\,\,.
\end{equation}
where $\Omega$, $R(\dot h)$ and $M(\dot h)$ depend parametrically on $v_P$. Close to the stability threshold of Eq. (\ref{linear-oscillator}) is sufficient to evaluate all terms in Eq. (\ref{oscill}) to leading order of an expansion in

\begin{equation}\label{epsilon}
\varepsilon\equiv v_P-v_C\,\,\,,\,\,\,F_P'(v_C)\equiv 0.
\end{equation}
Since, according to Eq. (\ref{linear-oscillator}), the linear part of $R(\dot h)$ in $\dot h$ is of order $\varepsilon$, the leading expression for Eq. (\ref{oscill}) has the general structure

\begin{equation}\label{osci}
\ddot h+\Omega^2(v_C)h=\varepsilon X_1(v_C)\dot h+Z(h,\dot h;v_C)\,\,.
\end{equation} 

If we, finally, apply the scaling transformation
 
\begin{equation}\label{scaling}
h(t)\equiv \varepsilon\,f(t)\,\,,
\end{equation}
we obtain, up to second order in $\varepsilon$, the representation  

\begin{equation}\label{standard}
\ddot f+\Omega^2f=\sum_{\nu=1,2}\varepsilon^\nu Q_\nu(f,\dot f)\,\,,
\end{equation}
which is the starting point of the work by Bogoliubov and Mitropolsky \cite{Bogoliubov} where in the present case

\begin{eqnarray}\label{Q}
Q_1(f,\dot f)&=&X_1\dot f+X_2{\dot f}^2+Y_1f\dot f\,\,,\\
Q_2(f,\dot f)&=&X_3\dot f^3+Y_2{f\dot f}^2\,\,.\nonumber
\end{eqnarray}

Following Ref. \cite{Bogoliubov}, we look for solutions of Eq. (\ref{standard}) in form of the expansion

\begin{equation}\label{ansatz}
f(t)=\alpha(t)\cos{\psi(t)}+\sum_{\nu=1,2}\varepsilon^\nu u_\nu(\alpha(t),\psi(t))\,\,,
\end{equation}
complemented by the constraints

\begin{eqnarray}\label{orthogonal}
&&\int_0^{2\pi}d\psi\,u_\nu(\alpha,\psi)\sin{\psi}=0\\
&&\int_0^{2\pi}d\psi\,u_\nu(\alpha,\psi)\cos{\psi}=0\,\,,\nonumber
\end{eqnarray}
which ensure that the $u_\nu(\alpha,\psi)$ for $\nu=1,2$ only contain higher harmonics in $\psi$. The equations

\begin{eqnarray}\label{a,psi}
&&\frac{d\alpha}{dt}=\sum_{\nu=1,2}\varepsilon^\nu A_\nu(\alpha)\,\,,\\
&&\frac{d\psi}{dt}=\Omega+\sum_{\nu=1,2}\varepsilon^\nu B_\nu(\alpha)\,\,,\nonumber
\end{eqnarray}
also assumed in Ref. \cite{Bogoliubov}, reflect the conditions that the amplitude $\alpha(t)$ and the difference $\psi(t)-\Omega\,t$ are slowly varying variables.

We are mainly interested in the functions $A_\nu(\alpha)$ and $B_\nu(\alpha)$, which can be obtained by projecting Eq. (\ref{standard}) onto the first harmonics $\sin{\psi}$ and $\cos{\psi}$. Then, due to Eqs. (\ref{orthogonal}), the contributions $u_\nu(\alpha,\psi)$ in Eq. (\ref{ansatz}) cancel, which allows us to look from the beginning for solutions of the simplified form 

\begin{equation}\label{f}
f(t)=\alpha(t)\cos{\psi(t)}\,\,.
\end{equation}
Its first and second derivatives are given by

\begin{eqnarray}\label{df/dt}
\dot f(t)&=&-\,\Omega\,\alpha(t)\sin{\psi(t)}\\&+&\sum_{\nu=1,2}\varepsilon^\nu(A_\nu\cos{\psi}-\alpha B_\nu\sin{\psi})\,\,,\nonumber
\end{eqnarray}

\begin{eqnarray}\label{d^3f/dt^2}
\ddot f(t)&=&-\,\Omega^2\,\alpha(t)\cos{\psi(t)}\\&-&2\Omega\,\sum_{\nu=1,2}\varepsilon^\nu
(A_\nu\sin{\psi}+\alpha B_\nu\cos{\psi})\nonumber\\&+&\varepsilon^2\left[\left(A_1\frac{dA_1}{d\alpha}-
\alpha B_1^2\right)\cos{\psi}\right.\nonumber\\&\,&\,\,\,\,\,\,\,\,-\left.\left(2A_1B_1+
\alpha A_1\frac{dB_1}{d\alpha}\right) \sin{\psi}\right]\,\,.
\nonumber
\end{eqnarray}
From this and the relations (\ref{Q}) we, finally, obtain

\begin{eqnarray}\label{A1}
&&B_1(\alpha)=-\,\frac{1}{4\pi\Omega}\int_0^{2\pi}Q_1(f,\dot f))\cos{\psi}=0\,\,,\\
&&A_1(\alpha)=-\,\frac{1}{4\pi\Omega}\int_0^{2\pi}Q_1(f,\dot f))\sin{\psi}=\frac{1}{4}X_1\alpha\,\,,\nonumber\\
&&A_2(\alpha)=-\,\frac{1}{4\pi\Omega}\int_0^{2\pi}Q_2(f,\dot f))\sin{\psi}=\frac{3}{16}X_3\alpha^3\,\,.\nonumber
\end{eqnarray}
After scaling back to the variable $h(t)$ via Eq. (\ref{scaling}), one recovers the result (\ref{amplitude}), where

\begin{equation}\label{alpha-a}
a(t)\equiv\varepsilon\,\alpha(t)\,\,.
\end{equation}
The coefficients $r_1,\rho_3$, appearing in this equation, depend on the parameters $\gamma,p,m$, and on the critical velocity $v_C$, and can be calculated from Eqs. (\ref{osci}), (\ref{coefficients}), and (\ref{Q}). For the choice $\gamma=0.01,p=100,m=0.003$, and $v_C=0.5214$ one finds $r_1=1.5575\cdot 10^{-2}$ and $\rho_3=9.885\cdot 10^{-5}$.

A. L. Korzhenevskii wants to express his gratitude to the University of D\"usseldorf for its warm hospitality. This work has been supported by the DFG under BA 944/3-3, and by the RFBR under N10-02-91332.

\end{document}